\documentclass[aip,preprint]{revtex4-1}
\draft
\usepackage{graphicx}
\usepackage{array}
\usepackage[dvipsnames]{xcolor}
\usepackage{tikz}
\usepackage{epstopdf, epsfig}
\usepackage{amsmath}
\usepackage{subcaption}
\usepackage{todonotes}
\usepackage{soul}
\usepackage{booktabs}
\usepackage{graphicx}
\usepackage{dcolumn}
\usepackage{bm}

\usepackage[utf8]{inputenc}
\usepackage[T1]{fontenc}
\usepackage{mathptmx}
\usepackage{etoolbox}
\usepackage{kotex}
\usepackage{CJKutf8}
\usepackage[dvipsnames]{xcolor}

\makeatletter
\def\@email#1#2{%
 \endgroup
 \patchcmd{\titleblock@produce}
  {\frontmatter@RRAPformat}
  {\frontmatter@RRAPformat{\produce@RRAP{*#1\href{mailto:#2}{#2}}}\frontmatter@RRAPformat}
  {}{}
}%
\makeatother

\begin{document}

\title{Data-driven discovery of drag-inducing elements on a rough surface through convolutional neural networks}

\author{Heesoo Shin (신희수)}
\affiliation{Department of Mechanical Engineering, Inha University, Incheon 22212, Republic of Korea}
\author{Seyed Morteza Habibi Khorasani}
\affiliation{FLOW, Department of Engineering Mechanics, KTH, Stockholm SE-100 44, Sweden}
\author{Zhaoyu Shi (史召宇)}
\affiliation{FLOW, Department of Engineering Mechanics, KTH, Stockholm SE-100 44, Sweden}
\author{Jiasheng Yang (楊佳晟)}
\affiliation{Institute of Fluid Mechanics, Karlsruhe Institute of Technology, Karlsruhe 76131, Germany}
\author{Shervin Bagheri}
\affiliation{FLOW, Department of Engineering Mechanics, KTH, Stockholm SE-100 44, Sweden}

\author{Sangseung Lee (이상승)}%
 \email{sangseunglee@inha.ac.kr}
\affiliation{Department of Mechanical Engineering, Inha University, Incheon 22212, Republic of Korea}

\date{\today}

\begin{abstract}
Understanding the influence of surface roughness on drag forces remains a significant challenge in fluid dynamics. This paper presents a convolutional neural network (CNN) that predicts drag solely by the topography of rough surfaces and is capable of discovering spatial patterns linked to drag-inducing structures. A CNN model was developed to analyze spatial information from the topography of a rough surface and predict the roughness function, $\Delta U^+$, obtained from direct numerical simulation. This model enables the prediction of drag from rough surface data alone, which was not possible with previous methods owing to the large number of surface-derived parameters. Additionally, the retention of spatial information by the model enables the creation of a feature map that accentuates critical areas for drag prediction on rough surfaces. By interpreting the feature maps, we show that the developed CNN model is able to discover spatial patterns associated with drag distributions across rough surfaces, even without a direct training on drag distribution data. The analysis of the feature map indicates that, even without flow field information, the CNN model extracts the importance of the flow-directional slope and height of roughness elements as key factors in inducing pressure drag. This study demonstrates that CNN-based drag prediction is grounded in physical principles of fluid dynamics, underscoring the utility of CNNs in both predicting and understanding drag on rough surfaces.
\end{abstract}


\maketitle

\section{\label{intro}Introduction}
The interaction between surface roughness and fluid flow is critical, particularly in scenarios involving turbulent flows. In these flows, the roughness elements of the surface can influence the smallest eddies near the wall, often resulting in increased drag. As heightened drag impedes the optimal functioning of various systems, including turbines, vehicles, and pipelines, the precise prediction of turbulent drag on rough surfaces is crucial.
While the increased drag from roughness can be determined from direct numerical simulations (DNSs) or experiments (e.g. towing tanks), these methods are not sustainable for \textit{drag prediction} \citep{chung2021predicting}. Indeed, capturing the full nonlinear interaction between irregular roughness structures and turbulent flows may not be necessary for an accurate and reliable drag prediction.

The majority of prior studies, including those by \citet{chan2015systematic}, \citet{thakkar2017surface}, \citet{forooghi2017toward}, and \citet{flack2020skin}, developed empirical relations that establish a correlation between drag and statistical surface parameters, such as skewness, effective slope and mean roughness height. Although these models present a good fit of the empirical data which they were developed for, they often poorly predict the drag of surfaces of different roughness types. 
The topographic data of rough surfaces, typically represented by two-dimensional height-map, could not be directly used in the empirical models owing to their large size. Instead, the surfaces had to be parameterized by statistical means, which does not capture all the spatial details of the surface topography.
In addition, statistical  parameterization complicates the identification of structural patterns on rough surfaces relevant to drag and reduces the physical interpretability of the predictive model. 

A second aspect is related to the modelling approach itself, since the accuracy of predictions depends significantly on capacity of the models. 
Recent developments have used artificial neural networks (ANNs), which are adept at managing complex, nonlinear problems. Studies by \citet{jouybari2021data} and \citet{lee2022predicting} have demonstrated the potential of fully connected networks (FCNs) in predicting drag on rough surfaces. Nonetheless, these models still depend on the statistical parameters of surface topography, which limits their ability to use the actual rough surface as an input. Consequently, this study aims to overcome these obstacles by directly employing unprocessed original topography for drag prediction using convolutional neural networks (CNNs). 
Given their ability to recognize spatial patterns, CNNs have been applied in various fluid engineering research areas\cite{lee2019data, murata2020nonlinear, morimoto2021convolutional, santos2020poreflow, deo2022predicting, jeon2022finite, jeon2024residual}.

In this study, we developed a CNN model trained on a DNS dataset of flows over both isotropic and anisotropic rough surfaces when $Re_{\tau}=500$ to predict the roughness function ($\Delta U^+$). This function represents the difference in the mean velocity profiles of the entire surface domain between smooth- and rough-wall turbulence within the log layer~\citep{hama1954boundary, clauser1954turbulent}. We observed that the prediction of the roughness function is based on the physical mechanism by which drag is induced on rough surfaces. Thanks to the model architecture designed to capture diverse feature scales and generate the feature map at the final layer, the developed CNN model, although trained to predict the scalar value $\Delta U^+$ without information about turbulent flow, can produce feature maps that closely resemble the drag force distribution obtained through DNS. These three-dimensional matrix outputs, resulting from nonlinear operations on rough surface inputs, illustrate the correlation between surface topography and drag. By comparing these outputs with drag-force distribution maps obtained from DNS and examining different aspects of rough surface topography, we have ascertained that the CNN model predicts $\Delta U^+$ by focusing on specific topographical features of surfaces that significantly influence the pressure drag.

The remainder of this paper is structured as follows: Section~\ref{data} describes the dataset of rough surfaces acquired through DNS, and Section~\ref{architecture} outlines the architecture of our CNN model. Section~\ref{result} presents the training outcomes and analyzes the ability of the CNN to identify spatial patterns associated with drag distributions, along with its limitations. Section~\ref{conclusion} provides concluding observations and directions for future studies.

\section{Rough surface dataset}\label{data}
This section describes the methodologies employed to create rough surface topographies and their corresponding DNS datasets, which are essential for training CNNs. It consists of two subsections: (1) the creation of rough surfaces, and (2) the computational specifics of DNS.

\subsection{Generation of rough surface topographies}\label{rough_surfaces}
The rough surfaces created in this study are classified based on the topographic metrics of skewness ($skw$) and effective slope for both streamwise ($ES_x$) and spanwise ($ES_z$) orientations. 
The definitions of these topographic indicators are as follows:
\begin{equation}
skw = \frac{1}{A_t} \int_{x,z} (k - k_{\text{avg}})^3 dA/k_{\text{rms}}^3,
\end{equation}

\begin{equation}
ES_x = \frac{1}{A_t}{\int_{x,z}{} \bigg|\frac{\partial k}{\partial x}\bigg| dA},
\end{equation}

\begin{equation}
ES_z = \frac{1}{A_t}{\int_{x,z}{} \bigg|\frac{\partial k}{\partial z}\bigg| dA}.
\end{equation}
Here, $x$, $y$, and $z$ represent the streamwise, wall-normal, and spanwise directions, respectively. Moreover, $k$ is the distribution of roughness height and $A_t$ denotes the total roughness plan area. Finally, $k_{\text{avg}}$ and $k_{\text{rms}}$ are defined as ${A^{-1}_t} \int_{x,z} k \, dA$ and $\sqrt{{A^{-1}_t} \int_{x,z} (k - k_{\text{avg}})^2 dA}$, respectively.

\begin{figure*}
\centerline{
\includegraphics[width=\textwidth]{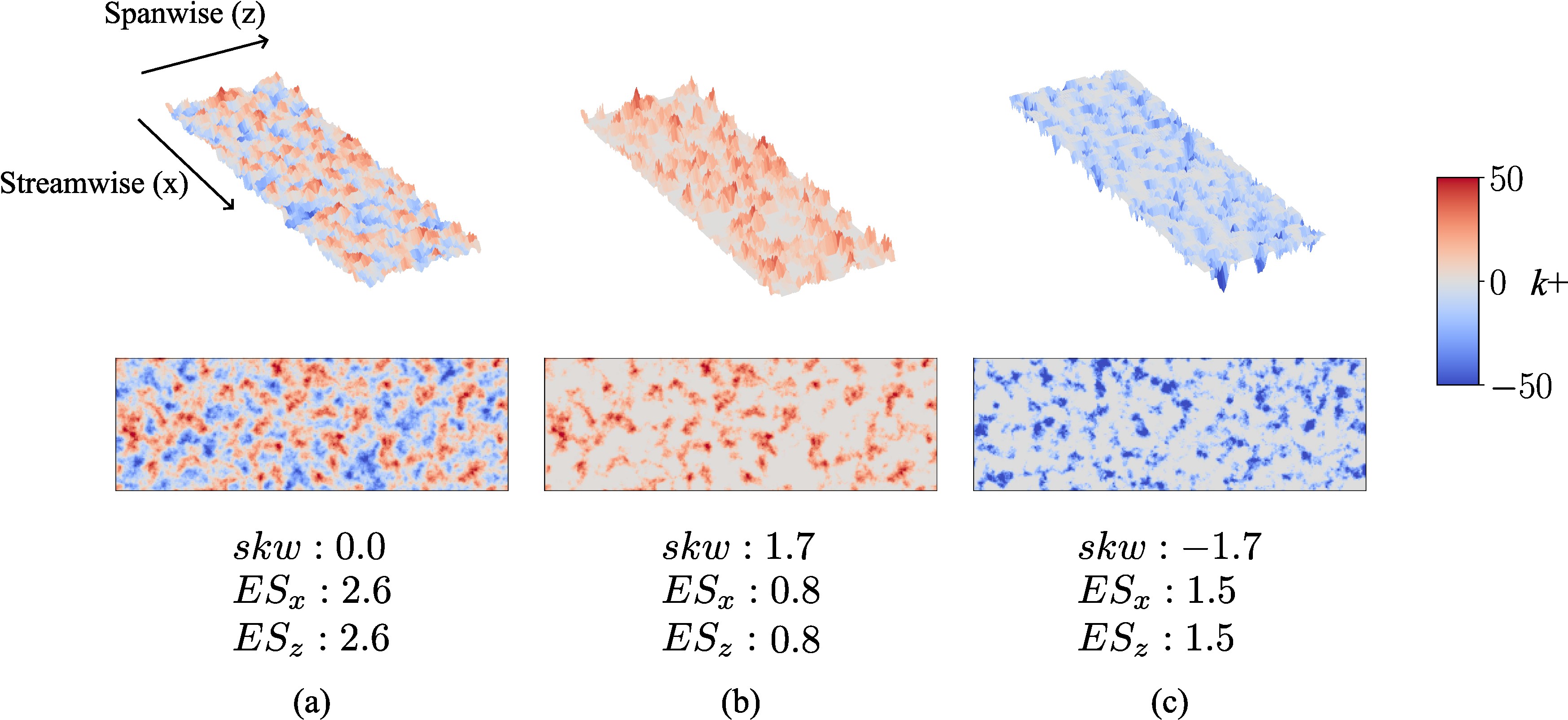}
}
\caption{Visualization of (a) $S_{Gauss}$, (b) $S_{pos}$, and (c) $S_{neg}$ in  3D (upper row) and 2D (lower row).}
\label{fig:iso_visual}
\end{figure*}

First, isotropic rough surfaces were generated featuring approximately equal values of $ES_x$ and $ES_z$. Three  categories of isotropic surfaces were generated: (i) Gaussian (zero-$skw$), (ii) positive-$skw$, and (iii) negative-$skw$ rough surfaces. Isotropic surfaces were generated using the Fourier-filtering algorithm and the code developed by \citet{jacobs2017quantitative}. First, Gaussian rough surfaces ($S_{Gauss}$) were produced by specifying a Gaussian probability distribution of $k$, resulting in an evenly balanced distribution of peaks and valleys with a $skw$ value of zero. By removing the heights below $k_{avg}$, we obtained positively skewed surfaces ($S_{pos}$) characterized by planes and peaks. Conversely, negatively skewed surfaces ($S_{neg}$) were generated by using the opposite method to $S_{pos}$, and are defined by planes and pits.
Therefore, these surfaces are distinguished based on their $skw$ value, which reflects the asymmetry in the distribution of $k$. In Fig.~\ref{fig:iso_visual}, the $S_{Gauss}$ sample exhibits a skewness ($skw$) of zero, while the $S_{pos}$ and $S_{neg}$ samples show positive and negative values, respectively. It is worth noting that by varying the roughness skewness, we constrain our observations to focus on isotropic roughness, as indicated by the identical $ES_x$ and $ES_z$ values.

\begin{figure*}
\centerline{
\includegraphics[width=0.7\textwidth]{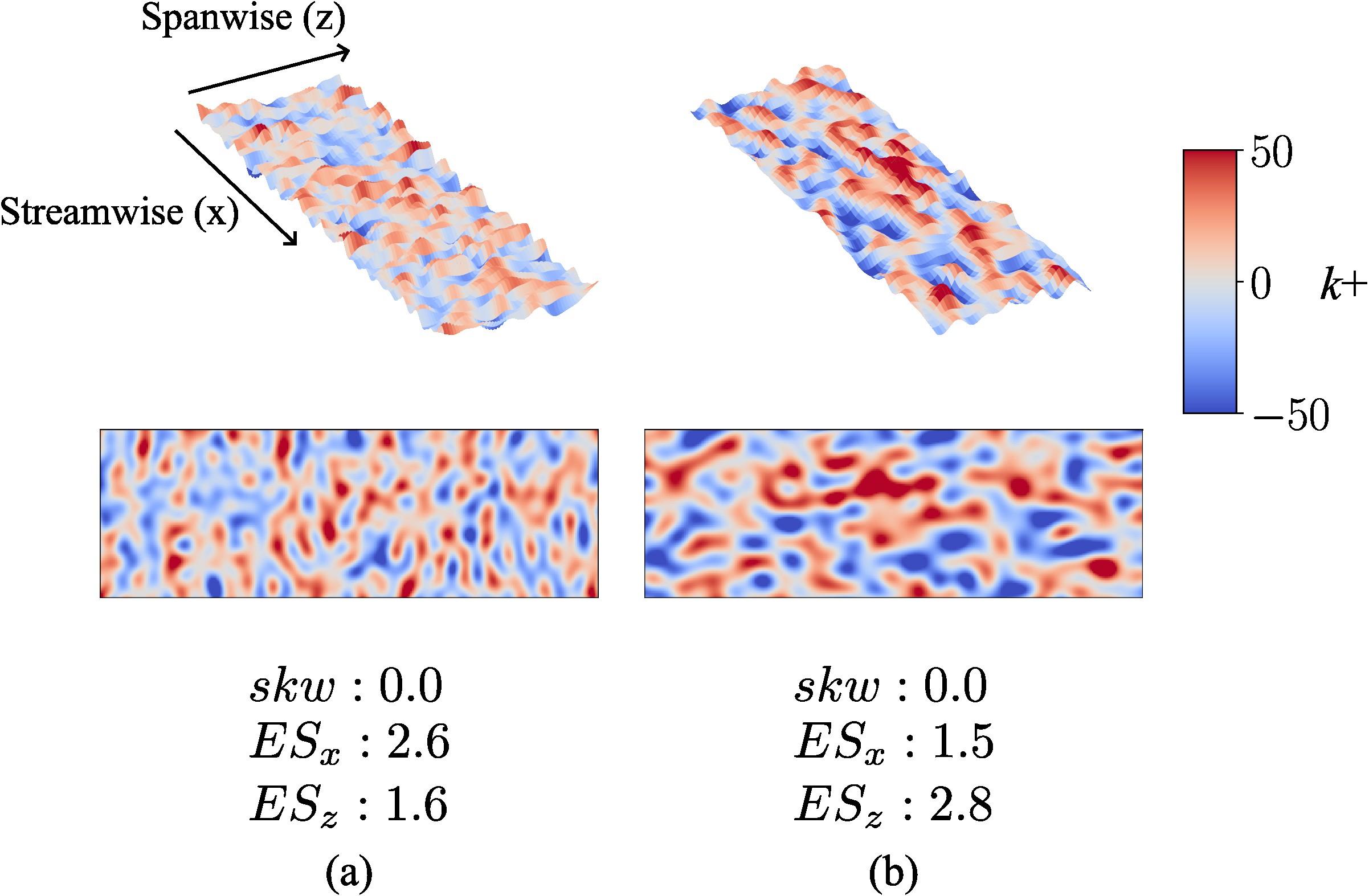}
}
\caption{Visualization of (a) $S_{ES_x}$ and (b) $S_{ES_z}$ in  3D (upper row) and 2D (lower row).}
\label{fig:aniso_visual}
\end{figure*}

Second, we generated anisotropic rough surfaces, which -- unlike isotropic surfaces -- exhibit directionality, leading to  differences between $ES_x$ and $ES_z$. 
To create these surfaces, we employed the multiscale anisotropic rough surface algorithm developed by~\citet{jelly2019multi}. This generation algorithm employs linear combinations of Gaussian random number matrices via a moving average process. Wherein, the wall-parallel correlation of the discrete heights is controlled through a target correlation function. By controlling the cut-off wavenumber for the circular Fourier filter, the number of streamwise points in the correlation function, and the number of spanwise points in the correlation function, we produced (i) $ES_x$-dominant anisotropic rough surfaces ($S_{ES_x}$) and (ii) $ES_z$-dominant anisotropic rough surfaces ($S_{ES_z}$). In Fig.~\ref{fig:aniso_visual}, the $S_{ES_x}$ sample shows larger $ES_x$ values compared with $ES_z$, leading to wave-like patterns in the streamwise direction. Conversely, the $S_{ES_z}$ sample has larger $ES_z$ values than $ES_x$, resulting in wave-like patterns in the spanwise direction. The skewness ($skw$) of both surfaces is zero.

Although the surfaces lack periodic boundary conditions, we utilized statistically converged rough surfaces to address this issue. In addtiion, \citet{chung2015fast} and \citet{macdonald2017minimal} shown that rough surfaces with structures below specific scales do not significantly affect first-order statistics, including drag. These findings imply that small-scale roughness mismatches do not have a substantial impact on key statistical measures. Accordingly, a criteria for domain size and roughness scales is defined by these authors as follows:
\begin{align}
L_x &\geq \max(1000\delta_{\nu}, 3L_z, \lambda_{r,x}), \\
L_y &\geq \frac{k_c}{0.15}, \\
L_z &\geq \max(100\delta_{\nu}, \frac{k_c}{0.4}, \lambda_{r,z}),
\end{align}
where $\delta_{\nu} = \nu/u_\tau$ and $\lambda_{r,x_i}$ represents the characteristic wavelength in the $x_i$ direction.
As a result, these rough surfaces can effectively capture crucial near-wall dynamics and accurately compute wall friction, mean velocities, and Reynolds stresses up to $y \approx 0.3\delta$, despite the small discontinuities exist at the domain ends.

Each surface type consists of 135 surfaces, which were doubled through the augmentation method described in Appendix~\ref{app:data}, yielding 270 surfaces. In addition, we added 162 hydrodynamically smooth surfaces, as detailed in Appendix~\ref{app:data}. The surfaces were subsequently divided into datasets: $60\%$ for training, $20\%$ for validation, and $20\%$ for testing, as detailed in Appendix~\ref{app:data}.

\subsection{Direct numerical simulations}
We developed a dataset of rough-wall turbulence in both the transitionally and fully rough regimes by solving the incompressible Navier-Stokes equations, as follows:
\begin{equation}
    \nabla \cdot \mathbf{u} = 0
    \label{eqn:continuity}
\end{equation}
\begin{equation}
    \frac{\partial \mathbf{u}}{\partial t} + \nabla \cdot (\mathbf{u}\mathbf{u}) = -\frac{1}{\rho} \nabla p + \nu \nabla^2 \mathbf{u} - \frac{1}{\rho}P_x \mathbf{e_x} + \mathbf{f_{\text{IBM}}}
    \label{eqn:NS_eq}
\end{equation}
where $\mathbf{u} = (u,v,w)$ is the velocity vector, and $P_x$ is the mean pressure gradient and $\mathbf{e_x}$ is the streamwise unit vector. $P_x$ is added as a constant source term to the momentum equation to drive the flow in the channel and accurately achieve $Re_{\tau}=500$. $p$ denotes the pressure fluctuations, $\mathbf{e_x}$ denotes the streamwise basis vector, $\rho$ denotes the density (set to 1 in this study), $\nu$ denotes the kinematic viscosity, and $\mathbf{f_{\text{IBM}}}$ denotes the IBM force that is added to enforce the no-slip and no-penetration conditions on the rough surfaces as done by \citet{kajishima2001turbulence} and \citet{Breugem2014}.
At every time step, $\mathbf{f_{\text{IBM}}}$ for a given grid cell with index ($i,j,k$) is calculated in the following manner:
    \begin{equation}
        \mathbf{f_{{\text{IBM}}_{\emph{i},\emph{j},\emph{k}}}} = \alpha_{i,j,k}\frac{\left(\mathbf{u}_s - \mathbf{u}^*\right)_{i,j,k}}{\Delta t},
    \end{equation}
where $\alpha_{i,j,k}$ is the solid volume fraction of the grid cell, $\mathbf{u}_s$ is the velocity of the solid interface, and $\mathbf{u}^*$ is the first prediction velocity obtained by integrating Eq. \eqref{eqn:NS_eq} in time. Since the rough surfaces do not move $\mathbf{u}_s = 0$ and a second prediction velocity accounting for the presence of the rough surface is obtained as:
    \begin{equation}
        \mathbf{u}^{**}_{i,j,k}=(1-\alpha_{i,j,k})\mathbf{u}^*_{i,j,k}.
    \end{equation}
This second prediction velocity is then corrected using the correction pressure of the fractional-step algorithm to give a final divergence-free velocity field.
The solver used is the open-source GPU-accelerated code CaNS \citep{costa2021gpu}. This solver is spatially second-order accurate and employs a fast Poisson solver, and temporally integrates the Navier-Stokes equations using a three-step Runge--Kutta scheme as part of a fractional-step algorithm \citep{KIM1985}. We adopted a minimal channel approach to minimize computational expenses while ensuring accurate results for $\Delta U^+$ \citep{chung2015fast}.
The simulations in the minimal-channel rough-wall DNS were conducted at a fixed friction Reynolds number $Re_{\tau}=u_\tau\delta/\nu=500$, where $\delta$ represents the channel half-height, and $u_{\tau}$ represents the friction velocity. Periodic boundary conditions were used along the $x$- and $z$-directions, with a Dirichlet boundary condition along the $y$-direction.
The domain dimensions $L_x$, $L_z$, and $L_y$ were set to $2.4\delta$, $2.0\delta$, and $0.8\delta$, respectively.
The numbers of grid points in the $x$- ($N_x$) and $z$- ($N_z$) directions were fixed at 302 and 102, respectively, with the grid spacings in the $x$- and $z$-directions in the viscous scale ($\Delta x^+$ and $\Delta z^+$) being 4.192 and 4.137, respectively. The superscript $+$ indicates normalization by the viscous scale $\delta_\nu = \nu/u_{\tau}$. In the $y$-direction, the grid was stretched using the hyperbolic tangent function with a minimum $y^+$ value $\approx$ 0.5. These grid sizes are confirmed by the grid convergence tests described in Appendix~\ref{app:DNS_validation} to ensure sufficient accuracy.  

The DNS results were used to obtain $\Delta U^+$, following the methodology described by~\citet{yang2022direct}. 
The logarithmic mean velocity profile, $U^+$, is expressed as:
\begin{equation}
U^+_R = \frac{1}{\kappa} \ln(y^+) + A + \Delta U^+ = U^+_S + \Delta U^+,
\label{eqn:U^+}
\end{equation}
where $\kappa \approx 0.4$ is the von Kármán constant and $A\approx5.0$ is log-law intercept for a smooth surface. The subscripts S and R denote smooth and rough surface quantities, respectively. The function $\Delta U^+$ depends on both the roughness topography and the roughness size, $k^+$. The latter is defined as:
\begin{equation}
k^+ = \frac{k}{\delta_{\nu}} = \frac{k u_{\tau}}{\nu}.
\label{eqn:k_plus}
\end{equation}

\begin{figure}
    \includegraphics[width=0.5\textwidth]{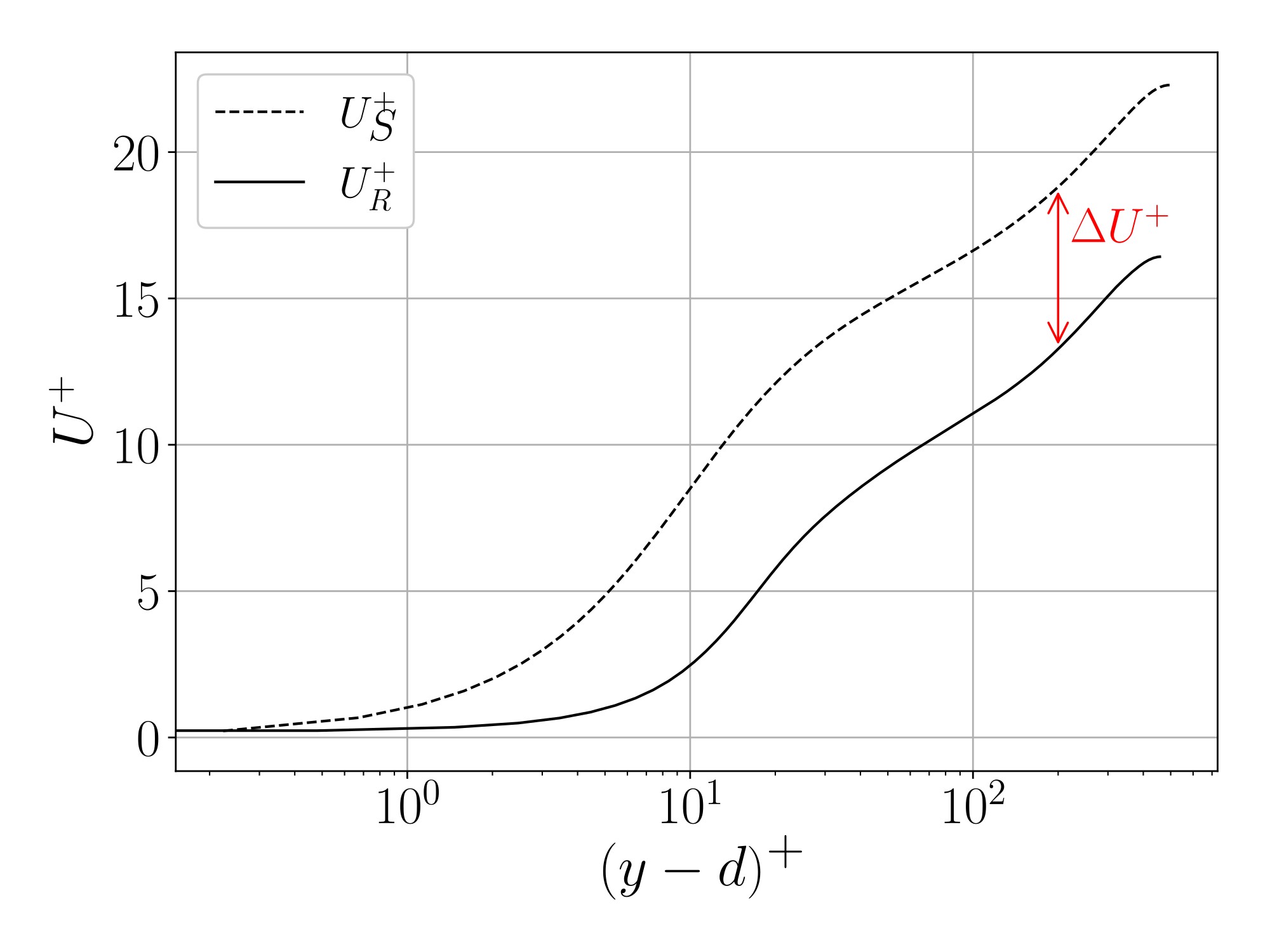}
    \caption{\label{fig:Uplus}$U^+_{S}$, $U^+_{R}$, and $\Delta U^+$ of the $S_{Gauss}$ sample.}
\end{figure}

We determine $\Delta U^+ = U^+_R - U^+_S$ at a designated reference point of $(y-d)^+ = 200$, as depicted in Fig.~\ref{fig:Uplus}, where $d$ indicates the zero-plane displacement of log-layer over rough surface, according to Jackson's definition of the centroid of the drag acting on the roughness \citet{jackson1981displacement}. Therefore, $\Delta U^+$ achieves a fixed value.
A positive $\Delta U^+$ indicates a momentum loss attributable to surface roughness, whereas a negative value indicates a momentum gain. Thus, $\Delta U^+$ serves as an indicator for drag resulting from surface roughness. Accordingly, our CNN model was trained to predict $\Delta U^+$.

Using DNS, we investigated the relationship between the topographical characteristics of rough surfaces and $\Delta U^+$ across different surface types.
\begin{figure*}
\centering
\includegraphics[width=1\linewidth]{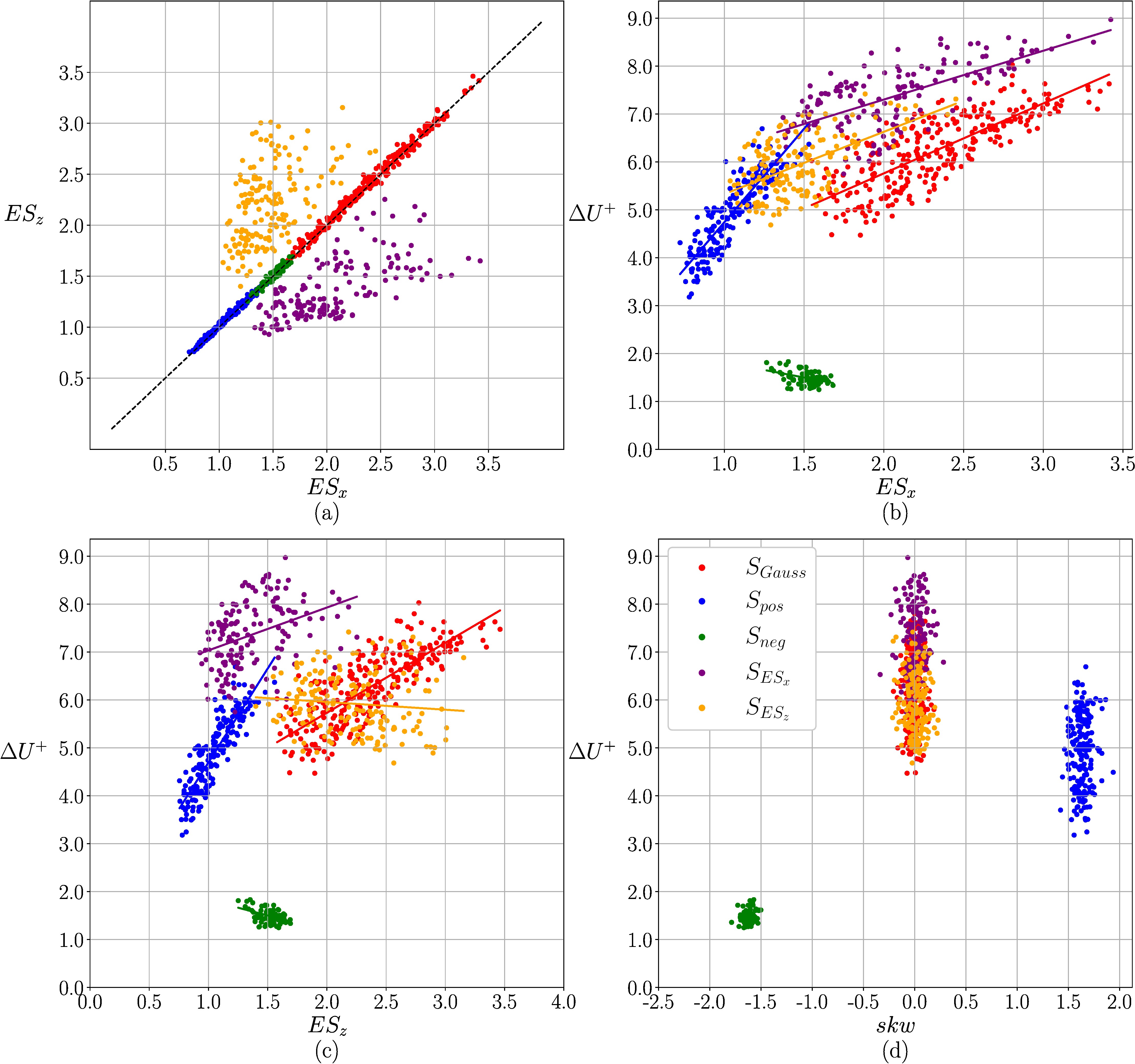}
\caption{(a) Correlation between $ES_x$ and $ES_z$ for different rough surfaces. (b) Correlation between $ES_x$ and $\Delta U^+$ for each surface. (c) Correlation between $ES_z$ and $\Delta U^+$ for each surface; the linear regression for each cluster is denoted by its lines in (b) and (c). (d) Correlation between $skw$ and $\Delta U^+$ for different rough surfaces}
\label{fig:total_features}
\end{figure*}
Fig.~\ref{fig:total_features}(a) shows the distribution of each surface as a function of $ES_x$ and $ES_z$. It is clear that the isotropic surfaces ($S_{Gauss}$, $S_{pos}$, $S_{neg}$) discussed earlier are distributed on the line $ES_x$ = $ES_z$, while anisotropic surfaces deviate significantly from the line due to their directionality. The scatter of the anisotropic data thus indicates the degree of surface anisotropy, which is controlled by the prescribed spatial separations in the streamwise and spanwise correlations, $\Delta x_i^*$, where the correlation function profiles reduce to 10\%. For more details, readers are referred to the original work by \citet{jelly2022impact}.
Fig.~\ref{fig:total_features}(b) shows the distribution of $\Delta U^+$ relative to the $ES_x$ values of the surfaces, revealing a positive correlation between an increase in $ES_x$ and a corresponding rise in $\Delta U^+$. A notable exception is $S_{neg}$.
The mechanism of drag generation on the $S_{neg}$ surface is distinct from that on other surfaces; consequently, we analyze the case of $S_{neg}$ separately in a section~\ref{limit}.
Fig.~\ref{fig:total_features}(c) shows that $ES_z$ is positively correlated with $\Delta U^+$ for $S_{pos}$, $S_{Gauss}$, and $S_{ES_x}$, mirroring the trend observed between $ES_x$ and $\Delta U^+$. However, variations in $ES_z$ do not significantly affect $\Delta U^+$. 
This is attributed to the fact that an increase in $ES_z$ typically results in an increase in $ES_x$ for $S_{Gauss}$, $S_{pos}$, and $S_{ES_x}$, but not for $S_{ES_z}$. In the case of $S_{neg}$, an increase in $ES_z$ exerts a negligible influence. 
Thus, these findings underscore that an enhancement in $ES_x$ generally leads to an increase in $\Delta U^+$. This relationship is due to $ES_x$ being directly proportional to twice the value of frontal solidity ($\lambda_f$), a measure reflecting the area exposed to pressure drag~\citep{chung2021predicting}. 
Finally, Fig.~\ref{fig:total_features}(d) illustrates the distribution of $\Delta U^+$ relative to the $skw$ values of the surfaces. Since $S_{pos}$ and $S_{neg}$ are modifications of $S_{Gauss}$ to have positive or negative $skw$ values, they exhibit distinct distributions of skewness values, as shown in this figure. Notably, $S_{neg}$ shows significantly lower $\Delta U^+$ values, ranging between 1 and 2, in contrast to the typical range of 3--8 observed for most surfaces.
This indicates that a significant number of generated surfaces fall within the fully-rough regime ($\Delta U^+ > 5.5$)~\cite{flack2010review, flack2020skin}. In this regime, the drag can be predicted using the information of the equivalent sand-grain roughness height, which is independent of the Reynolds number. However, it is important to note that this Reynolds number independence does not apply to $S_{neg}$, as their $\Delta U^+$ values are relatively small, resulting in a small equivalent sand-grain roughness height. Therefore, the Reynolds number effect could be significant in the data for $S_{neg}$.

\begin{figure}
\includegraphics[width=0.8\textwidth]{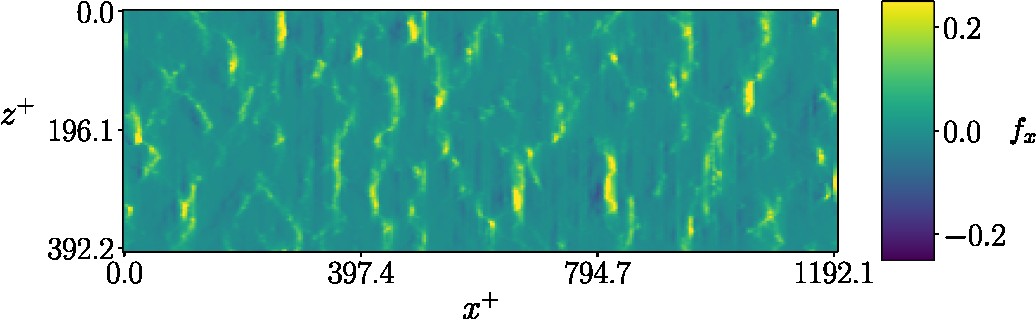}
\caption{\label{fig:dns_ex}Visualization of $f_x$ corresponding to the rough surface in Fig.~\ref{fig:iso_visual} (a).}
\end{figure}

Our aim is not only to predict $\Delta U^+$ but also to demonstrate that our model trained to predict $\Delta U^+$ learns the dominant drag-inducing mechanisms. We extracted drag maps from DNS to provide a detailed visual representation of drag force distribution on the rough surface. Fig.~\ref{fig:dns_ex} shows an example of a DNS-derived drag map, $f_x$,  where $f_x$ is the streamwise component of the wall-integrated mean IBM force,
\begin{equation}
(f_x, f_y, f_z) = -\frac{1}{T} \int_0^H \int_0^T \mathbf{f}_\text{IBM}(x,y,z,t) dt dy.
\label{eqn:IBM_force}
\end{equation}
The negative sign indicates that the forces act in the opposite direction of the flow. 
An overall increase in the magnitude of $f_x$ signifies a loss of the streamwise momentum, correlating with an increase in $\Delta U^+$. From in Fig.~\ref{fig:dns_ex}, we note that the regions where the roughness significantly contribute to drag are spanwise elongated.

These DNS drag maps provide a means for both quantitative and qualitative evaluations in comparison with the feature maps produced by the CNN model. This comparison enhances our understanding of the effectiveness of the model and the physical phenomena it encapsulates. To assess the ability of the CNN to accurately reflect the physics across different rough surface types, we obtained DNS drag maps for three samples from each rough surface category, which were not used in training the model ($S_{Gauss,i}$, $S_{pos,i}$, $S_{neg,i}$, $S_{ES_x,i}$, and $S_{ES_z,i}$, where $i=1,2,3$).

\section{\label{architecture}CNN architecture} 
In contrast to FCNs used in previous studies~\citep{lee2022predicting, jouybari2021data}, CNNs can process high-dimensional data for training without omitting the spatial information contained within, owing to convolutional operations. In this study, we developed a deep neural network based on the CNN framework to preserve the spatial information of rough surfaces and directly utilize it as input.

\begin{figure*}
\centerline{
\includegraphics[width=0.78\textwidth]{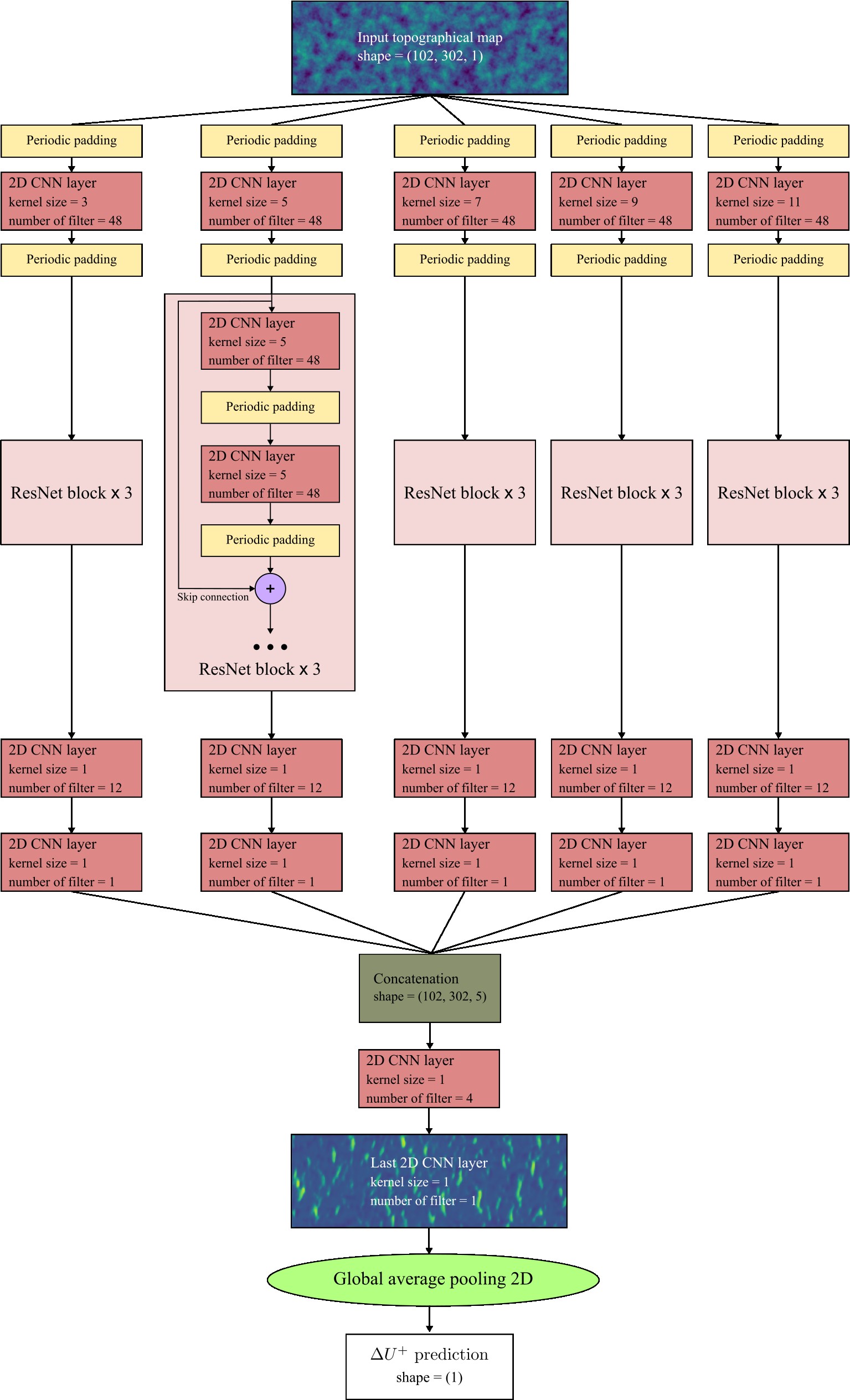}} 
\caption{\label{fig:cnn_architecture}Illustration of CNN architecture utilized in this study.}
\end{figure*}

Fig.~\ref{fig:cnn_architecture} shows the detailed CNN model architecture used in this study. Regarding the structure of the model, this section will focus on two architectural elements introduced specifically to deal with rough surfaces: periodic boundary conditions and a parallel structure. For additional information on other structural features of the model, refer to Appendix~\ref{app:architecture}. Moreover, the CNN model was trained using training and validation datasets, and its hyperparameters were finetuned through Bayesian optimization, as detailed in Appendix~\ref{app:opt}.

\begin{figure}
\includegraphics[width=0.5\textwidth]{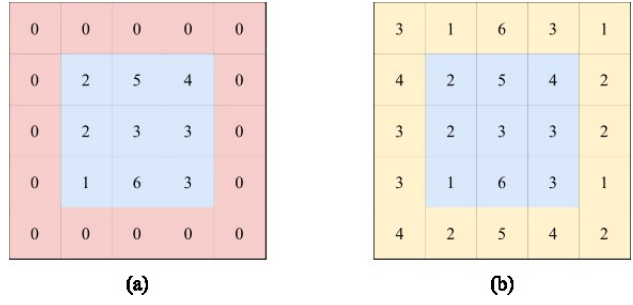}
\caption{\label{fig:PDC2D}(a) and (b) Original feature map in blue: (a) zero padding areas (in red), (b) domain expanded via periodic padding (in yellow).}
\end{figure}

Since zero values exist at the edges of the surface data due to the traditional zero-padding method, which differs from the DNS condition, we used periodic boundary padding to preserve dimensionality during convolution operations and to mimic the periodic boundary conditions observed in the DNS, specifically along the $x$- and $z$-directions. This technique expands the input feature map along the $x$- and $z$-axes, thus maintaining the cyclic nature of the boundaries in alignment with the DNS. Fig.~\ref{fig:PDC2D} shows a comparison between this padding approach and the traditional zero-padding method using an example.

Given that rough surfaces consist of roughness elements of various scales, these elements need to be considered in predicting drag. In this context, the parallel structure of our CNN is designed to detect roughness elements at various scales. The changes in the CNN feature map according to the kernel size are detailed in Appendix~\ref{app:architecture}. Applying the inception module introduced by~\citet{szegedy2015going}, our CNN model utilizes a range of kernel sizes from $3 \times 3$ to $11 \times 11$, corresponding to the grid sizes. This diversity facilitates the detection of surface features at multiple scales. As depicted in Fig.~\ref{fig:cnn_architecture}, toward the end of the CNN, feature maps from different kernels within the parallel structure were merged. This was followed by a convolution with a $1 \times 1$ kernel employing a single filter, a process that combines the features while maintaining the original input dimensions. Subsequently, a comprehensive CNN feature map was produced. This map was subject to global average pooling (GAP), generating a scalar value representative of $\Delta U^+$.

\section{\label{result}Evaluation of prediction performance and physics learnability} 
Next, we evaluate the trained CNN model from two perspectives: its accuracy in predicting $\Delta U^+$, and its ability to capture the mechanisms of drag induced on rough surfaces. The latter entails generating feature maps that that can be gauged against the drag force distributions of the DNS drag maps. As the topographical, CNN feature, and DNS drag maps depict distributions of $k^+$, $\Delta U^+$, and $f_x$, respectively, we standardized each map using the following equation:
\begin{equation}
    \tilde{\mathbf{m}} = \frac{\mathbf{m} - \mu}{\sigma},
    \label{eqn:std_scaler}
\end{equation}
where $\tilde{\mathbf{m}}$ represents the standardized form of a given map ($\mathbf{m}$), which is a two-dimensional matrix, $\mu$ is the mean of $\mathbf{m}$, and $\sigma$ is the standard deviation of $\mathbf{m}$. Lastly, Huber loss ($L_H$) is used as the training loss of the model, as devised by \citet{huber1992robust}. It combines mean squared error and mean absolute error. The formula for $L_H$ is given by: 
\begin{equation} 
L_{H}(y, y') = 
\begin{cases} 
\frac{1}{2} (y - y')^2 & \text{for } |y - y'| \leq \delta, \\
\delta \left(|y - y'| - \frac{1}{2} \delta \right) & \text{for } |y - y'| > \delta 
\end{cases} 
\end{equation} \
where $y$ is the ground truth $\Delta U^+$, $y'$ is the predicted $\Delta U^+$, $x$ is the input, and $\delta = 1.0$ is a threshold parameter that determines the sensitivity to outliers.

\subsection{$\Delta U^+$ prediction performance of the CNN model} \label{CNN_eval}
The evaluation of the predictive accuracy of the model for $\Delta U^+$ is quantified using the mean absolute error (MAE) and the coefficient of determination ($R^2$). The MAE is defined as follows:
\begin{equation}
\text{MAE} = \frac{1}{N} \sum\limits_{i=1}^{N} \left| y_i - \tilde{y}_i \right|,
\label{eqn:MAE}
\end{equation}
where $N$ represents the number of samples in the test dataset, $y_i$ is the actual $\Delta U^+$, and $\tilde{y}_i$ is the $\Delta U^+$ predicted by the CNN model. The $R^2$ metric is defined as
\begin{equation}
R^2 = 1 - \frac{\sum\limits_{i=1}^{N}(y_i - \tilde{y}_i)^2}{\sum\limits_{i=1}^{N}(y_i - \overline{y})^2},
\label{eqn:R2}
\end{equation}
where $\overline{y}$ is the average of the actual $\Delta U^+$.
Lower MAE values indicate an improved predictive accuracy. An $R^2$ value close to 1 signifies high precision, whereas a value close to 0 indicates lower reliability. 
\begin{ruledtabular}
\begin{table}
    \begin{center}
 \def~{\hphantom{0}}
    \begin{tabular}{lccc}
        & \text{MAE} & \text{MAPE(\%)} & \text{Maximum absolute error} \\
        \hline
       $S_{Gauss}$ &  0.111 & 1.797  & 0.407 \\
       $S_{pos}$ & 0.128 &  2.698 & 0.372 \\
       $S_{neg}$ & 0.062 & 4.075  & 0.229 \\
       $S_{ES_x}$ & 0.149 & 2.431  & 0.579 \\
       $S_{ES_z}$ & 0.145 & 1.961  & 0.433 \\
    \end{tabular}
    \caption{ MAE, MAPE, and maximum absolute error of the $\Delta U^+$ prediction by the CNN model for each surface type in the test dataset.}
    \label{tab:mae_mape}
    \end{center}
\end{table}
\end{ruledtabular}
The average prediction accuracy across all the surface types was 0.106 in terms of the MAE and 0.996 in terms of $R^2$. Additionally, the prediction accuracy for each type of surface is detailed in Table~\ref{tab:mae_mape}. In this table, we used the mean absolute percentage error (MAPE) to indicate the accuracy for each surface type. The MAPE is defined as
\begin{equation}
\text{MAPE} (\%) = \frac{1}{N} \sum\limits_{i=1}^{N} \frac{\left| y_i - \tilde{y}_i \right|}{y_i} \times 100,
\label{eqn:MAPE}
\end{equation}
According to the Table~\ref{tab:mae_mape}, all the surface types demonstrated a comparable prediction accuracy.
However, $S_{neg}$ showed larger errors compared with the other types of surfaces. The reasons for the larger errors specifically in predicting drag on $S_{neg}$ are discussed in Section~\ref{limit}. The ability of the CNN model to capture drag generation factors for various surface types and generate feature maps resembling their drag distribution is validated and analyzed in Sections~\ref{assesment} to \ref{feature_based_analysis}.

\subsection{Assessment of physics learnability through feature map analysis}
\label{assesment}
This section evaluates the ability of the CNN model to capture the main physics of drag inducement across four different surface types ($S_{Gauss}$, $S_{pos}$, $S_{ES_x}$, and $S_{ES_z}$). These surface type demonstrate a comparable predictive performance (Table~\ref{tab:mae_mape}). As DNS drag maps originate from solutions to the Navier--Stokes equations, a CNN feature map that closely resembles the DNS drag map suggests that the CNN model effectively captures the mechanisms of drag generation on rough surfaces. Therefore, we evaluated the similarity between the CNN feature maps and the DNS drag maps.

To evaluate the similarity between the CNN feature maps and the DNS drag maps, or between the CNN feature maps and the topographical maps, we calculated the root mean squared error (RMSE) and structural similarity index measure (SSIM). The RMSE is defined as
\begin{equation}
\text{RMSE} = \sqrt{\frac{1}{N_x N_z} \sum_{i=1}^{N_{x}} \sum_{j=1}^{N_{z}} (a_{i,j} - b_{i,j})^2},
\label{eqn:RMSE}
\end{equation}
where $a_{i,j}$ and $b_{i,j}$ represent the pixels of arbitrary maps $\mathbf{a}$ and $\mathbf{b}$ with $N_x \times N_z$ dimension.
These could be any two among a CNN map, a DNS drag map, or a topographical map.

The SSIM was originally a method for comparing the similarity between two images, devised by~\citet{wang2004image}. The method evaluates the similarity based on three components of the image: luminance ($l$), contrast ($c$), and structure ($s$). In this study, these components are interpreted as follows: (i) $l$ represents regions of higher or lower map values, (ii) $c$ denotes areas with significant variations in map values, and (iii) $s$ evaluates the spatial arrangement of map values, corresponding to the organization of patterns across the map. The SSIM is defined as:
\begin{equation}
    l(\mathbf{a}, \mathbf{b}) = \frac{2\mu_{\mathbf{a}} \mu_{\mathbf{b}} + c_{1}}{\mu_{\mathbf{a}}^2 + \mu_{\mathbf{b}}^2 + c_{1}}, \label{eqn:luminance}
    \end{equation}
    \begin{equation}
    c(\mathbf{a}, \mathbf{b}) = \frac{2\sigma_{\mathbf{a}} \sigma_{\mathbf{b}} + c_{2}}{\sigma_{\mathbf{a}}^2 + \sigma_{\mathbf{b}}^2 + c_{2}},
    \label{eqn:contrast}
    \end{equation}
    \begin{equation}
    s(\mathbf{a}, \mathbf{b}) = \frac{\sigma_{\mathbf{a} \mathbf{b}} + c_{3}}{\sigma_{\mathbf{a}} \sigma_{\mathbf{b}} + c_{3}}, \label{eqn:structure}
\end{equation}
where
\begin{align}
    \mu_{\mathbf{a}} &= \frac{1}{N_x N_z} \sum_{i=1}^{N_x} \sum_{j=1}^{N_z} a_{i,j}, \\
    \mu_{\mathbf{b}} &= \frac{1}{N_x N_z} \sum_{i=1}^{N_x} \sum_{j=1}^{N_z} b_{i,j}, \\
    \sigma_{\mathbf{a}}^2 &= \frac{1}{N_x N_z} \sum_{i=1}^{N_x} \sum_{j=1}^{N_z} (a_{i,j} - \mu_{\mathbf{a}})^2, \\
    \sigma_{\mathbf{b}}^2 &= \frac{1}{N_x N_z} \sum_{i=1}^{N_x} \sum_{j=1}^{N_z} (b_{i,j} - \mu_{\mathbf{b}})^2, \\
    \sigma_{\mathbf{a} \mathbf{b}} &= \frac{1}{N_x N_z(N_x N_z-1)} \sum_{i=1}^{N_x} \sum_{j=1}^{N_z} (a_{i,j}- \mu_{\mathbf{a}})(b_{i,j} - \mu_{\mathbf{b}}).
\end{align}
Here, $c_1=(k_1 L)^2$, $c_2=(k_2 L)^2$, and $c_3 = c_2 / 2$ with $k_1 = 0.01$ and $k_2 = 0.03$, and $L$ is defined as the difference between the maximum and minimum values among $\mathbf{a}$ and $\mathbf{b}$.
The overall SSIM index was computed as the product of these three components:
\begin{equation}
    \text{SSIM}(\mathbf{a}, \mathbf{b}) = l(\mathbf{a}, \mathbf{b})^\alpha \cdot c(\mathbf{a}, \mathbf{b})^\beta \cdot s(\mathbf{a}, \mathbf{b})^\gamma,
    \label{eqn:SSIM}
\end{equation}
where the weights $\alpha$, $\beta$, and $\gamma$ are all 1.
A SSIM value close to 1 indicates a high degree of similarity between the maps. This measure is crucial in evaluating how effectively the CNN model has captured and replicated the drag-inducing physics of the rough surfaces in the DNS.

\begin{figure*}
    \centering{
    \includegraphics[width=1.0\textwidth]{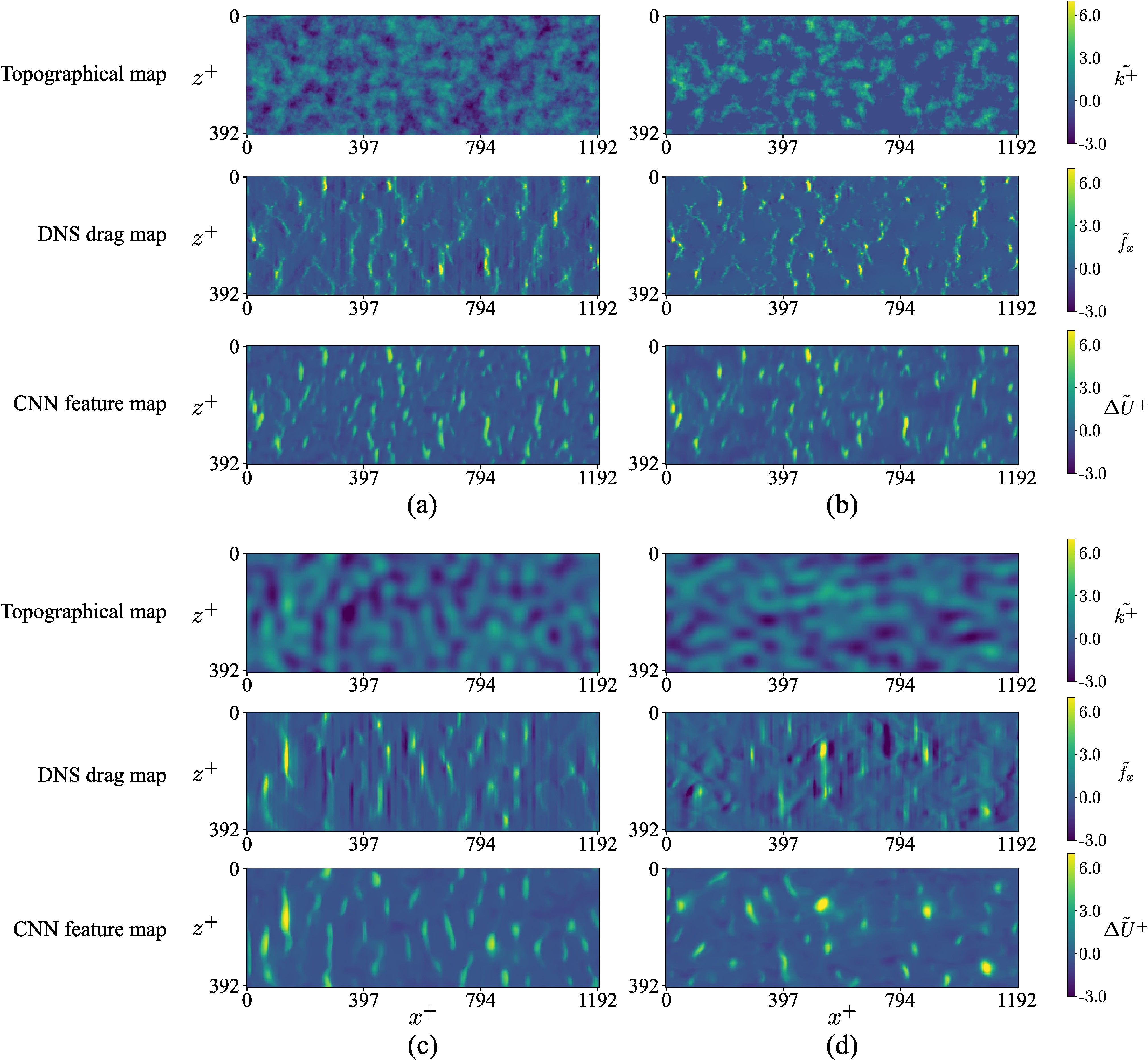}}
    \caption{\label{fig:SSIM_2d}Comparison of maps for each surface type: (a) $S_{Gauss, 1}$, (b) $S_{pos, 1}$, (c) $S_{ES_x, 1}$, and (d) $S_{ES_z, 1}$ }
\end{figure*}

\begin{ruledtabular}
\begin{table}
\begin{tabular}{lcccc}
& $S_{Gauss}$ & $S_{pos}$ & $S_{ES_x}$ & $S_{ES_z}$ \\
\hline
MAPE ($\%$) & 1.220 & 1.907 & 2.003 & 3.090 \\
Mean SSIM (CNN-DNS) & 0.490 & 0.594 & 0.474 & 0.410 \\
Mean RMSE (CNN-DNS) & 0.813 & 0.766 & 0.789 & 0.874 \\
Mean SSIM (CNN-Topo) & 0.148 & 0.411 & 0.229 & 0.173 \\
Mean RMSE (CNN-Topo) & 0.987 & 0.793 & 0.865 & 0.968 \\
\end{tabular}
\caption{MAPE of $\Delta U^+$ prediction, mean SSIM, and mean RMSE for each surface type. CNN-DNS refers to the comparison between the CNN feature map and the DNS drag map, while CNN-Topo refers to the comparison between the CNN feature map and the topographical map.}
\label{tab:CNN_APE_SSIM_RMSE}
\end{table}
\end{ruledtabular}

Fig.~\ref{fig:SSIM_2d} displays sample topographical, DNS drag, and CNN feature maps from $S_{Gauss}$, $S_{pos}$, $S_{ES_x}$, and $S_{ES_z}$. Additionally, to aid visual understanding, both the CNN feature map and the DNS drag map are visualized in 3D in Appendix~\ref{app:3d_vis}. The RMSE and SSIM values between these maps are listed in Table~\ref{tab:CNN_APE_SSIM_RMSE}. According to the table, the SSIM between the CNN feature maps and the DNS drag maps is higher than that between the CNN feature maps and the topographical maps, consistent with the lower RMSE. This indicates that the CNN feature maps resemble the DNS drag maps more closely than they do the topographical maps. Building on this assessment, we investigated the high-intensity patterns in the CNN feature maps and DNS drag maps. Fig.~\ref{fig:SSIM_2d} reveals elongated, spanwise high-intensity patterns in the CNN feature maps, similar to those in the DNS drag maps. The pattern is also similar to the area distribution of roughness elements facing the flow in the corresponding topographical map, particularly visible in Fig.~\ref{fig:SSIM_2d}(b). This provides significant evidence of the ability of the CNN to predict drag, considering that the mechanism of drag induction on rough surfaces and the effective slope in the $x$-direction predominantly influence the pressure drag. Additionally, it demonstrates that the CNN model learned to determine the flow direction without being provided with any flow-related information.

The DNS drag maps in Fig.~\ref{fig:SSIM_2d} and Fig.~\ref{fig:3d_maps} reveal a distinct concentration of $\tilde{f_x}$ on positive slopes when viewed in the direction of the flow. This is particularly evident for roughness elements with positive slopes and heights exceeding the mean surface level, where the pressure drag is more pronounced than the viscous drag~\citep{napoli2008effect}. Conversely, the drag distribution observed in the opposite direction shows lower concentrations, attributable to the reduced presence of $\tilde{f_x}$. Similarly, the CNN feature maps in Fig.~\ref{fig:SSIM_2d} and Fig~\ref{fig:3d_maps} effectively reflect this distribution, focusing on the wall-normal structures similar to the DNS drag maps. The model highlights the force disparity between the flow and counterflow directions. This alignment underscores the capability of our model to capture the topographical features critical to predicting $\Delta U^+$, demonstrating its ability to recognize spatial patterns in surface structures that predominantly induce pressure drag, even without information about the drag distribution or turbulent flow.

However, there are notable discrepancies in the drag force distributions on the planes of the CNN feature maps compared with those of the DNS drag maps. Generally, the plane areas of the CNN feature maps of Fig.~\ref{fig:SSIM_2d} and Fig.~\ref{fig:3d_maps} display lower values compared with those of the DNS drag maps. These divergences highlight the limitations of the CNN model in accurately predicting drag distributions for components beyond pressure drag.
\begin{figure*}
\centering
\includegraphics[width=0.8\textwidth]{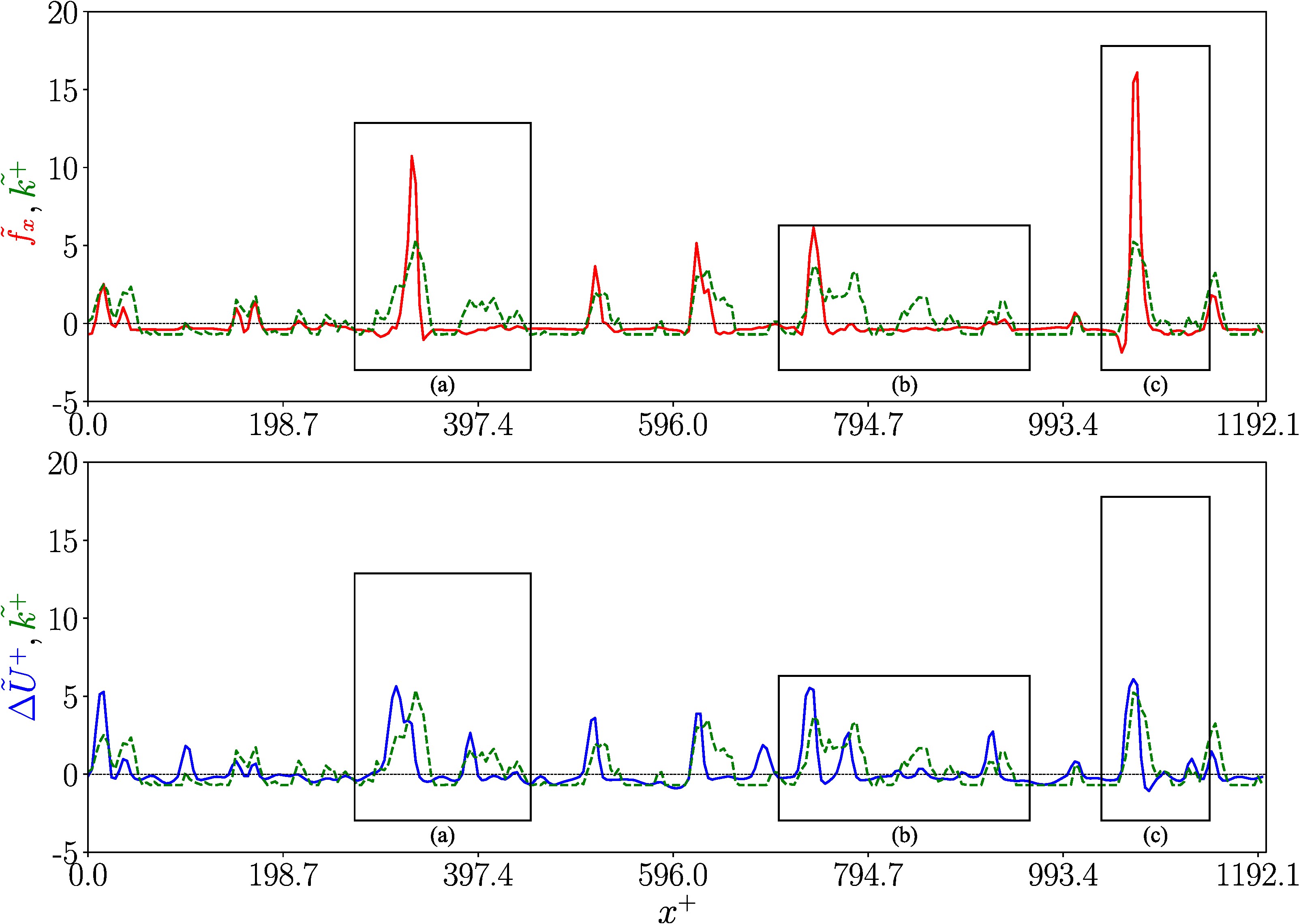}
\caption{DNS drag map and topographical map (upper row) and CNN feature map and topographical map (lower row) at $z^+ = 23.5$ for $S_{pos,2}$. Boxes (a), (b), and (c) highlight the zones where neighboring peak structures in the front create prominent shadowed areas behind them.}
\label{fig:shadow}
\end{figure*}
Additionally, the CNN model struggles to capture areas that are sheltered behind other roughness peaks and experience negligible local drag force, referred to as shadowed areas~\citep{yang2022direct}. Fig.~\ref{fig:shadow} shows the line graphs of the DNS drag map, CNN feature map, and topographical map at $z^+ = 23.5$, where the highest peak in the topographical map for $S_{pos,2}$ is located. Boxes (a), (b), and (c) in the figure. illustrate a reduction in local drag in the shadowed areas in the DNS drag map, whereas this reduction is either absent or less pronounced in the CNN feature map. This suggests that the model has not completely accounted for the physics associated with the shadowed areas.

In addition, we analyzed the three maps (topographical, CNN feature, and DNS drag maps) of each surface sample in the wavenumber domain. This analysis aims to compare the scale of the dominant spatial patterns in each map. First, we calculated the spanwise-averaged premultiplied power spectral density (PSD), denoted by $k_x^+ \Phi$. We assessed the similarity of $k_x^+ \Phi$ between the topographical, CNN feature, and DNS drag maps.
The similarity between two different $k_x^+ \Phi$ (i.e. $k_x^+ \Phi_{1}$ and $k_x^+ \Phi_{2}$) is quantified using the Euclidean distance (ED), calculated as follows:
\begin{equation}
ED = \sqrt{\sum^{N_x}_{i=0} \left [k_x^+ \Phi_{1, i} - k_x^+ \Phi_{2, i}\right ]^2}.
\label{eqn:ED}
\end{equation}
\newcommand{\redline}{\raisebox{2pt}{\tikz{\draw[-,Red,dashed,line width = 1pt](0,0) -- (10mm,0);}}}
\newcommand{\greenline}{\raisebox{2pt}{\tikz{\draw[-,Green,dashdotted,line width = 1pt](0,0) -- (10mm,0);}}}
\newcommand{\blueline}{\raisebox{2pt}{\tikz{\draw[-,blue,solid,line width = 1pt](0,0) -- (10mm,0);}}}
\begin{figure*}
\centering
\includegraphics[width=\textwidth]{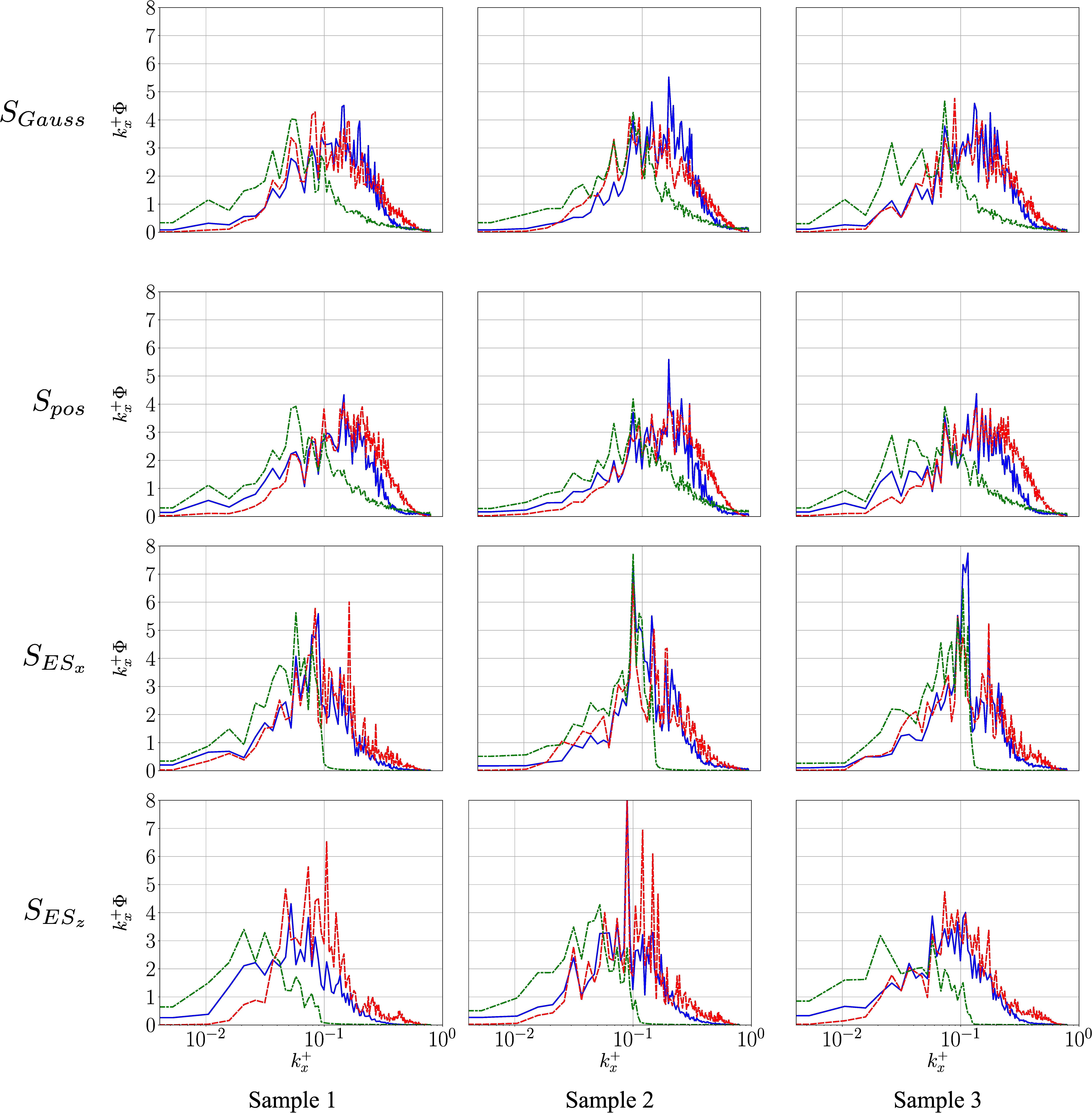}
\caption{Streamwise-averaged $k_x^+ \Phi$ profiles for each surface type, depicted using a topographical map (\protect\greenline), CNN feature map (\protect\blueline), and DNS drag map (\protect\redline).}
\label{fig:PSD_topo_dns_cnn}
\end{figure*}
\begin{ruledtabular}
\begin{table}
\begin{center}
\begin{tabular}{lcccccccccccc}
&$S_{Gauss}$ & $S_{pos}$ & $S_{ES_x}$ & $S_{ES_z}$ \\
\hline
CNN-DNS & 6.518 & 7.303 & 7.798 & 7.112 \\
CNN-Topo & 13.184 & 10.137 & 11.980 & 10.219 \\
\end{tabular}

\caption{Average ED of $k_x^+ \Phi$ between the DNS drag maps and the CNN feature maps (top) and the topographical maps and the CNN feature maps (bottom) of three samples.}
\label{tab:ED_comparison}
\end{center}
\end{table}
\end{ruledtabular}

Fig.~\ref{fig:PSD_topo_dns_cnn} shows the $k_x^+ \Phi$ lines of the topographical maps, DNS drag maps, and CNN feature maps, and Table~\ref{tab:ED_comparison} presents the calculated ED between the CNN feature map and both the DNS drag map and the topographical map for all the sampled surfaces. 
For the CNN feature map. $k_x^+ \Phi$ closely resembles that of the DNS drag map, indicating that the CNN feature map has similar dominant spatial patterns.
Furthermore, we analyzed the similarities in the peaks of $k_x^+ \Phi$. For example, in $S_{Gauss,1}$, the wavenumbers of the three primary peaks of $k_x^+ \Phi$, identified by their highest values, are as follows: 0.084, 0.157, and 0.099 for the DNS drag map; 0.147, 0.199, and 0.094 for the CNN feature map; and 0.052, 0.037, and 0.079 for the topographical map.
Subsequently, the $\lambda_x^+$ values are 75.0, 40.0, and 63.158 for the DNS drag map; 42.857, 31.579, and 66.667 for the CNN feature map; and 120, 171.429, and 80.0 for the topographical map.
This analysis suggests a closer alignment in peak distribution between the DNS drag map and the CNN feature map compared with that between the CNN feature map and the topographical map. Thus, the critical patterns in both the CNN feature map and the DNS drag map exhibit similar scales.

\subsection{Topographical-characteristics-based analysis}
\label{feature_based_analysis}
We conducted an analysis using topographical characteristics maps, distinct from CNN feature maps, to identify specific topographical characteristics captured by the CNN model for predicting $\Delta U^+$. These maps, comprising a range of topographical elements, were combined with distinct weights and subsequently visualized. This combination aimed to create a ``composite map" resembling the CNN feature map, thereby elucidating the topographical characteristics essential to the predictive accuracy of the model.

The foundation of this analysis lies in the topography-derived maps, which encompass both the roughness height and the surface gradient map. The composite map comprises six base maps: $T^t$, $T^b$, $T^m$, $G_x^t$, $G_x^m$, and $G_x^b$. Here, $T$ denotes the topographical map, and $G_x$, representing the gradient of $T$, is calculated as follows:
\begin{equation}
G_x = \frac{\partial T}{\partial x} \approx \frac{T_{\text{fw}} - T_{\text{bk}}}{2\Delta x},
\end{equation}
where $T_{\text{fw}}$ and $T_{\text{bk}}$ denote the forward and backward positions by one grid point on the input map, respectively. The superscripts ($t$, $b$, and $m$) for $T$ and $G_x$ differentiate the maps based on specific thresholds. For example, $T^t$ represents the values in the top $25\%$ of all $T$ values, $T^b$ identifies those in the bottom $25\%$, and $T^m$ includes the values between the bottom $25\%$ and the top $25\%$.

We combined these base maps to create a composite map. This process involved merging two base maps, resulting in nine additional combination maps: $T^t G_x^t$, $T^t G_x^b$, $T^t G_x^m$, $T^m G_x^t$, $T^m G_x^b$, $T^m G_x^m$, $T^b G_x^t$, $T^b G_x^b$, and $T^b G_x^m$. For instance, $T^t G_x^t$ represents the integration of the top $25\%$ of $T$ values with the top $25\%$ of gradient values from $G_x$. All base maps and combination maps were standardized using Eq.~\ref{eqn:std_scaler}.
Subsequently, we determined the optimal weights, $w_i$ (where $i = 1, 2, ..., 15$), to determine the most effective combination ratio that reflects the CNN feature map. This was achieved using the following equation:
\begin{equation}
\textbf{m}_c = w_1 \tilde{T^t} + w_2 \tilde{T^m} + w_3 \tilde{T^b} + ... + w_{15} \tilde{T^b G_x^b},
\end{equation}
where $m_c$ represents the composite map created using the topographical characteristics maps to resemble the CNN feature map.
We used stochastic gradient descent for the optimization of the weights ($w_1$ to $w_{15}$). This method iteratively refines the weights using the least squares method, which measures the difference between $\textbf{m}_c$ and the CNN feature maps. During each iteration, a subset of data is used to calculate the gradient of the loss function with respect to $\textbf{w}$, guiding the adjustments required to better align with the CNN feature map. The weight update follows the equation:
\begin{equation}
\textbf{w}_{\text{new}} = \textbf{w}_{\text{old}} - \eta \cdot \nabla_w L(\textbf{w}_{\text{old}}),
\end{equation}
where $\textbf{w}_{\text{new}}$ and $\textbf{w}_{\text{old}}$ are the updated and previous weight vectors, respectively, $\eta$ is the learning rate, and $\nabla_\textbf{w} L(\textbf{w}_{\text{old}})$ is the gradient of the loss function $L$ with respect to $\textbf{w}$ at the previous iteration.

\begin{figure*}
    \centering
    \includegraphics[width=\textwidth]{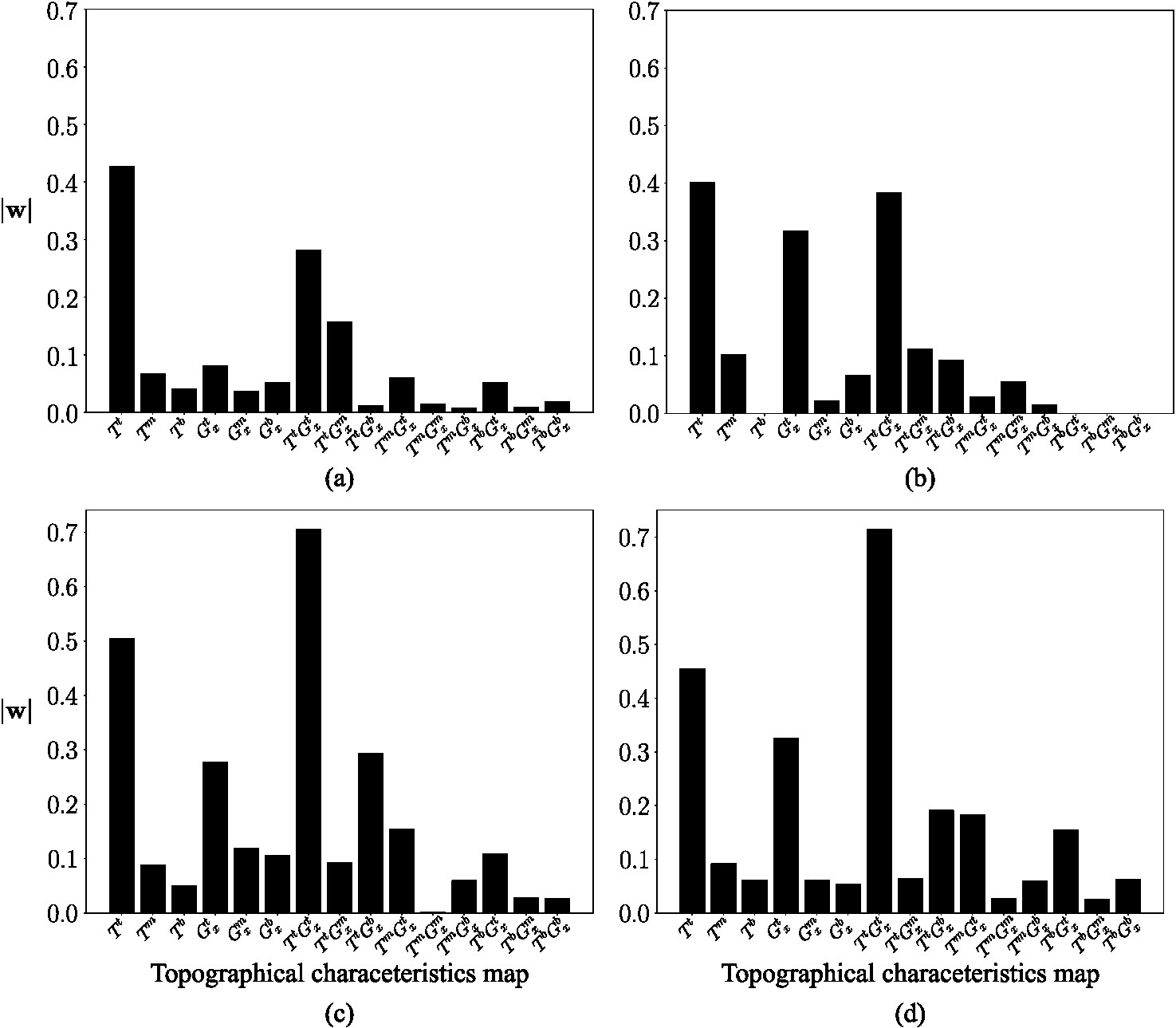}
    \caption{Optimized $|\textbf{w}|$ values of the composite map for each surface sample: (a) $S_{Gauss, 1}$, (b) $S_{pos, 1}$, (c) $S_{ES_x, 1}$, and (d) $S_{ES_z, 1}$.}
    \label{fig:topo_comp}
\end{figure*}
Fig.~\ref{fig:topo_comp} shows the optimized weights for each topographical characteristics map. According to this figure, the two most prominent surface features across various surface types are $T^t$ and $T^tG_x^t$. Given that high roughness elements and positive gradients in the streamwise direction significantly influence pressure drag, this analysis suggests that the CNN model predominantly focuses on the topographical elements of rough surfaces that induce pressure drag to predict $\Delta U^+$. This is consistent with the analysis in the previous sections, where high values were distributed in the counterflow areas of roughness elements.

Accordingly, the results discussed in this section corroborate the analyses presented in the preceding section~\ref{CNN_eval}, demonstrating that our model primarily focuses on the peaks of rough surfaces and the positive gradients of these peaks in the direction of fluid flow. Therefore, given the close correlation between pressure drag for roughness heights larger than the viscous sublayer and the frontal area of rough surfaces, our model focuses on the pressure drag on rough surfaces. This analysis aligns with the findings of previous studies \citep{aghaei2022contributions, chung2021predicting, orlandi2006turbulent}, which have shown that pressure drag significantly contributes to the total drag in a fully rough regime.  Additionally, the less precise distribution of CNN feature maps in planar areas of rough surfaces or regions below the mean height indicates the limitations of the model in addressing forces other than the pressure drag. Moreover, our model does not accurately represent the force weakening in the shadowed areas of rough surfaces. In the following section, we will explore drag prediction on surfaces with negative skewness, where the pressure drag is not the predominant factor, unlike on other surfaces analyzed in previous sections, to clarify the limitations of our model further.

\subsection{Limitations in predicting negative-$skw$ surfaces}\label{limit}
\begin{figure*}
    \centering
\includegraphics[width=0.85\textwidth]{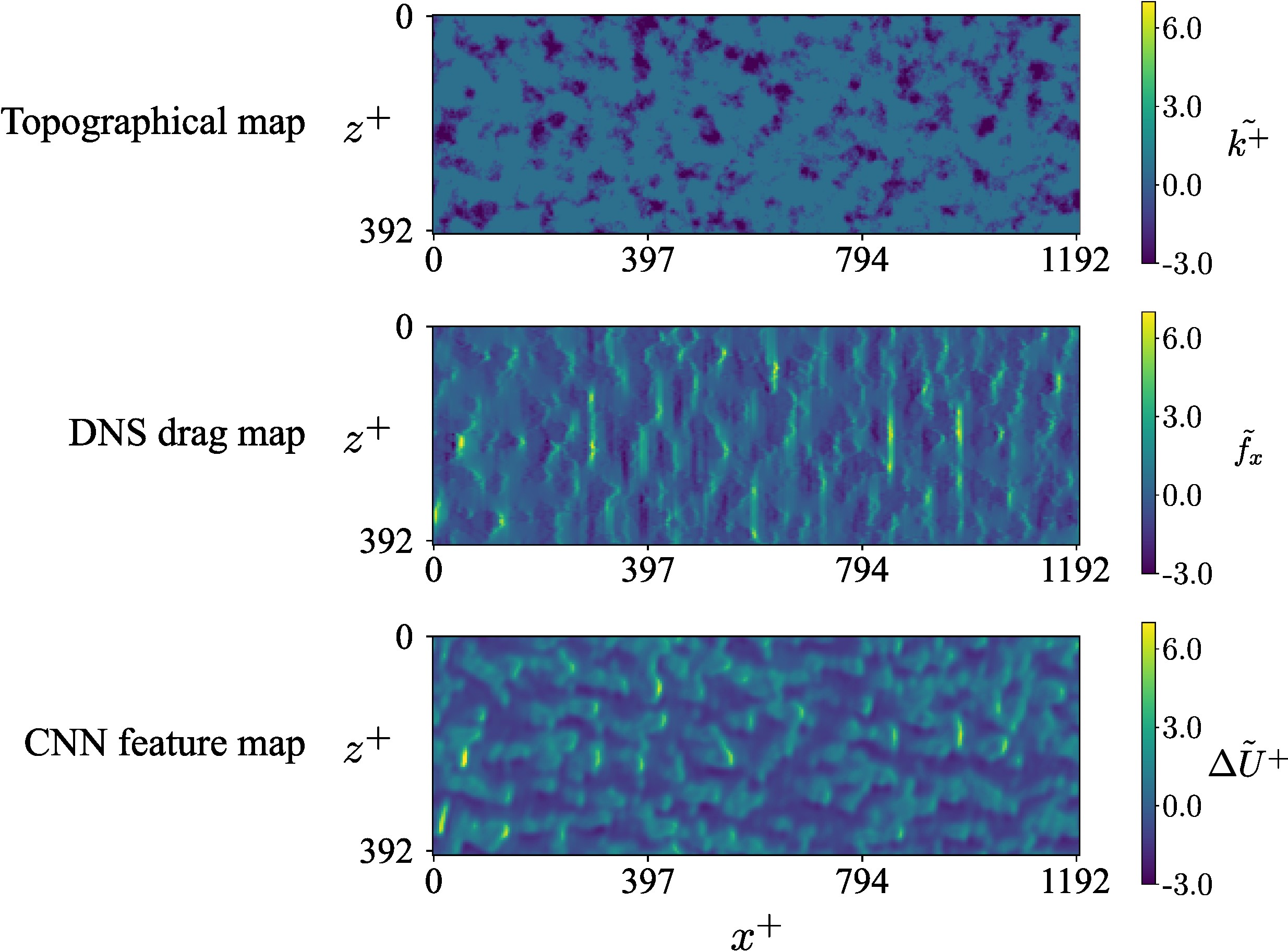}
    \caption{Visualization of the topographical, DNS drag, and CNN feature maps for $S_{neg,1}$.}
    \label{fig:neg_skewness_comparison}
\end{figure*}

Although our CNN model has proven effective in predicting $\Delta U^+$ based on the mechanism of the drag inducement for $S_{Gauss}$, $S_{pos}$, $S_{ES_x}$, and $S_{ES_z}$, it encounters challenges in accurately predicting $S_{neg}$ when compared with these surface types. As indicated in Table~\ref{tab:mae_mape}, the MAPE for $S_{neg}$ is almost twice as high as that for the other surfaces. Additionally, we validated the data using five datasets, each created by excluding one type of surface from the initial training dataset, which included all surface types.
\begin{ruledtabular}
\begin{table}
    \centering
    \begin{tabular}{lcccccc}
         & D1 & D2 & D3 & D4 & D5 & D6\\
         \hline
        MAE & 0.1084 & 0.5119 & 0.3226 & 0.2734 & 0.4357 & 0.3330 \\
        $R^2$ & 0.9964 & 0.9230 & 0.9758 & 0.9832 & 0.9563  & 0.9757 
    \end{tabular}
    \caption{MAE and $R^2$ of models trained on each dataset: D1 (original dataset), D2 (excluding $S_{pos}$), D3 (excluding $S_{neg}$), D4 (excluding $S_{Gauss}$), D5 (excluding $S_{ESx}$), and D6 (excluding $S_{ESz}$).}
    \label{tab:data_val}
\end{table}
\end{ruledtabular}
Table~\ref{tab:data_val} shows that excluding $S_{neg}$ from the dataset resulted in the lowest prediction accuracy. This indicates that $S_{neg}$ contains unique information that is not easily learned from other surface types, highlighting a significant distinction between $S_{neg}$ and the other surfaces.
Moreover, the similarity between the DNS drag maps and CNN feature maps for $S_{neg}$ is significantly lower than that for the other surface types (see Fig.~\ref{fig:neg_skewness_comparison}). Consequently, this section discusses these limitations and investigates the reasons behind the reduced prediction accuracy and limited physics learnability for $S_{neg}$.

In the DNS drag maps of Fig.~\ref{fig:neg_skewness_comparison} and Fig.~\ref{fig:neg_skewness_3d}, the pits in the DNS map exhibit lower values than the planes, whereas in the CNN feature map, the pits show higher values than the planes. This occurs because the model fails to capture accurately the dominant physics of turbulent drag on $S_{neg}$ surfaces, primarily focusing on regions with positive slopes and peaks while neglecting the plane and pit regions, similar to its performance in predicting the drag on $S_{Gauss}, S_{pos}, S_{ES_x}, S_{ES_z}$. Specifically, for $S_{neg}$, the average of the absolute SSIM between the sampled topographical maps and the CNN feature maps is 0.039, and that between the sampled DNS drag maps and the CNN feature maps is 0.058. These values are lower than the SSIM values between the CNN feature maps and the DNS drag maps compared with other surface types, as shown in Table~\ref{tab:CNN_APE_SSIM_RMSE}, where the average absolute SSIM for all types is 0.492.

\begin{figure*}
    \centering
\includegraphics[width=1\textwidth]{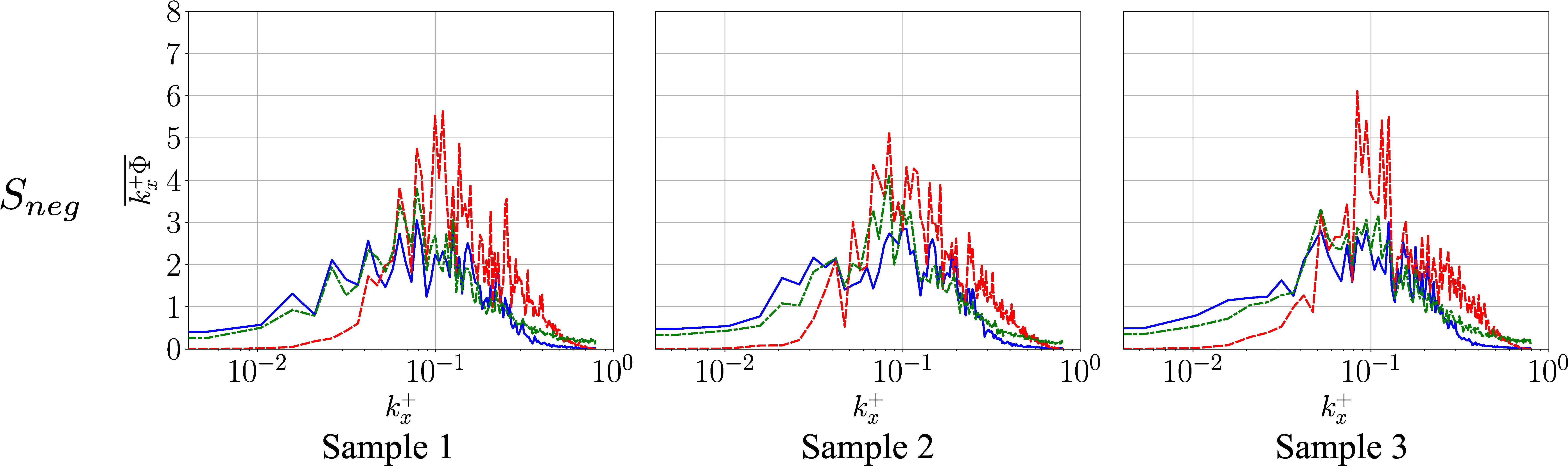}
    \caption{$k_x^+ \Phi$ profiles for $S_{neg}$, depicted using the topographical map (\protect\greenline), CNN feature map (\protect\blueline), and DNS drag map (\protect\redline).}
    \label{fig:psd_neg}
\end{figure*}

According to the wavenumber domain analysis, $S_{neg}$ shows a closer alignment of the CNN feature map with the topographical map than with the DNS drag map unlike other surface types (see Fig.~\ref{fig:psd_neg}). Specifically, the average ED between the sampled topographical map and the CNN feature map was 3.920, which is smaller than the ED between the sampled DNS drag map and the CNN feature map, which was 10.231. In other words, the scale of spatial patterns in the CNN feature map of $S_{neg}$ more closely resembles that in the topographical map, diverging from the distributions observed in the DNS drag map. We also observed the discrepancies between CNN and DNS on the $S_{neg}$ surface in the wavenumber domain in the two-dimensional premultiplied PSD. For more details, refer to Appendix~\ref{app:2d_psd}.

From an examination of previous studies, the inaccurate prediction of $S_{neg}$ can be attributed to several factors. According to \citet{busse2023effect}, the ratio of the pressure drag force ($F_p$) to the total drag force ($F_{tot}$), which includes both viscous and pressure drag forces, is higher for surfaces with positive or zero-$skw$ values than for those with negative-$skw$ values. Specifically, on surfaces with negative-$skw$ values, characterized by planar and pit features similar to our $S_{neg}$, the $F_p/F_{tot}$ ratio is reported to be smaller than 0.3. This suggests a relatively minor role of $F_p$ in $F_{tot}$ for $S_{neg}$ compared with the surfaces with zero or positive-$skw$ values.
The study by \citet{flack2020effect} explored the influence of skewness on the drag of rough surfaces. The authors discovered that peak-dominant surfaces, or rough surfaces with $skw$ values ranging from zero to positive, induce a higher drag compared with surfaces with negative-$skw$ values. They proposed that the lower drag induction on negative-$skw$ surfaces is due to flow skimming over surface depressions. Thus, the limited accuracy of the CNN model in predicting $S_{neg}$ surfaces, coupled with its diminished ability to capture physics beyond the pressure drag distribution, highlights its limited focus on learning the pressure drag when predicting $\Delta U^+$ and its reduced understanding of other factors influencing the total drag.
Furthermore, the complex flow phenomena inherent to $S_{neg}$, as described by \citet{jimenez2004turbulent}, contribute to the challenges of predicting and understanding the physics of $S_{neg}$. Surfaces with negative-$skw$ values, featuring grooves and classified as either $k$-type or $d$-type based on their groove aspect ratio, exhibit complexities owing to recirculation vortices affecting the logarithmic layer offset. These complexities highlight the limitations of both the CNN model and the dataset in comprehensively understanding and learning the intricate flow dynamics of $S_{neg}$.

\section{\label{conclusion}Conclusion} 
This study demonstrates the effectiveness of CNNs in predicting $\Delta U^+$, a critical parameter for evaluating turbulent drag on rough surfaces induced by fluid flow. The key findings of this study include the following:
\begin{enumerate}
    \item Our developed CNN model predicts $\Delta U^+$ using only the rough surface topography as input, eliminating the need for extracting surface parameters and manually selecting them.
        
    \item The feature map generated by our CNN model closely resembles the DNS drag maps, which indicate the distribution of drag generated by a flow over a rough surface. The model primarily identifies regions with high roughness elements. Additionally, it focuses on positive slopes in the wall-normal direction of roughness elements, which are linked with the frontal area of the rough surfaces and are strongly correlated with the pressure drag. Consequently, the CNN feature map exhibits elongated patterns in the spanwise direction, which are also observed in the DNS drag maps. Therefore, our model extracts the main mechanism of drag induced on rough surfaces with positive or zero skewness by learning the physical relationship between scalar $\Delta U^+$ and matrix surface topography, predicting drag based on this mechanism.

    \item Although our CNN model effectively predicts $\Delta U^+$ for surfaces where the pressure drag is dominant, it exhibits diminished predictive accuracy for surfaces with negative skewness, where the pressure drag is not the primary source of drag. Additionally, the CNN feature map shows reduced similarity to the DNS drag maps for surfaces with negative skewness. This underscores the limitations of the CNN model in capturing the drag mechanisms of surfaces where pressure drag is not the predominant factor. Additionally, flow phenomena other than pressure drag, such as the effects of shadowed areas or recirculation vortices in negatively skewed surface pits, are not well captured.
\end{enumerate}

Future studies should aim to improve the learnability of ANNs for complex physics of drag on rough surfaces to enable more robust and accurate drag predictions across different surface types. This objective requires expanding the training dataset to encompass a wider variety of rough surface patterns, such as $k$- and $d$-type negative skewness surfaces and nonhomogeneous rough surfaces such as those introduced by \citet{medjnoun2018characteristics}. Furthermore, given that the rough surfaces in this study consist exclusively of randomly arranged roughness elements, and noting that the model's performance deteriorates on untrained surface types, as discussed in Section~\ref{limit}, incorporating data on surfaces with regular arrangements into the dataset— as suggested by \citet{womack2022turbulent}—could improve the dataset's diversity and potentially enhance model performance.

Considering that complex flow phenomena like recirculation vortices are multidimensional, applying a 3D-CNN in the future could be more advantageous for learning these multidimensional phenomena. Additionally, utilizing transfer learning with results from empirical correlations, as proposed by \citet{lee2022predicting}, could reduce training time and enhance model performance. 
Since the current model lacks flow-related conditions, incorporating key parameters such as $Re_{\tau}$, along with expanding the dataset to include a range of $Re_{\tau}$ values that comprehensively cover the transitionally-rough regime, could improve the model ability to capture the precise physics involved.

Moreover, although this study primarily investigated surfaces increasing drag ($\Delta U^+$ $>$ 0), the inclusion of surfaces that decrease drag ($\Delta U^+$ $<$ 0) under certain conditions, such as riblets~\citep{bechert1997experiments, garcia2011drag}, could offer valuable insights into the dynamics of rough surfaces. This strategy could assist in designing surfaces to minimize drag, by employing machine-learning methods. 
Furthermore, based on this study, we aim to develop a prediction model capable of visualizing drag distribution and instantly predicting the flow drag generated on the surface by processing diverse images of rough surfaces.

\section*{Declaration of interests}{The authors report no conflict of interest.}

\begin{acknowledgments}
Institute of Energy Technology Evaluation and
Planning(KETEP) grant funded by the Korea Government(MOTIE) (RS-2023-00243974, Graduate
School of Digital-based Sustainable Energy Process Innovation Convergence), the National Research Foundation of Korea Grant funded by the Korean Government(NRF-2022R1F1A1066547), and the Inha University Research Grant. SMHK, ZS and SB acknowledge the National Academic Infrastructure for Supercomputing in Sweden (NAISS) and the Swedish Energy Agency for funding the research.
\end{acknowledgments}

\appendix
\section{Validation of the DNS solver} \label{app:DNS_validation}
\newcommand{\blackline}{\raisebox{2pt}{\tikz{\draw[-,black,solid,line width = 1pt](0,0) -- (10mm,0);}}}
\newcommand{\DNSblueline}{\raisebox{2pt}{\tikz{\draw[-,blue,dashdotted,line width = 1pt](0,0) -- (10mm,0);}}}
\newcommand{\DNSredline}{\raisebox{2pt}{\tikz{\draw[-,red,dashed,line width = 1pt](0,0) -- (10mm,0);}}}
We employed the same resolution criteria as those reported in \citet{jouybari2021data} and \citet{yuan2014estimation}. We validated the grid convergence of the DNS solver by halving and doubling the $y$ grid numbers used in this study.
\begin{ruledtabular}
\begin{table}
    \centering
    \begin{tabular}{lccc}
        &$\text{Base grid}$&$\text{Coarse grid}$&$\text{Fine grid}$ \\
        \hline
        $N_y$ & $312$ & $156$ & $770$ \\
        $\Delta y^+_{\text{min}} ; \Delta y^+_{\text{max}}$ & $1.051 ; 8.283$ & $1.051 ; 16.563$ & $0.422 ; 3.375$ \\
    \end{tabular}
    \caption{$y$-directional grid resolutions utilized in the grid convergence study. The Base grid represents the grid resolution employed in this study. The Coarse grid is a grid resolution half that of the Base grid, and the Fine grid is a grid resolution twice that of the Base grid.}
    \label{tab:GCS}
\end{table}
\end{ruledtabular}
\begin{figure}
\includegraphics[width=0.5\textwidth]{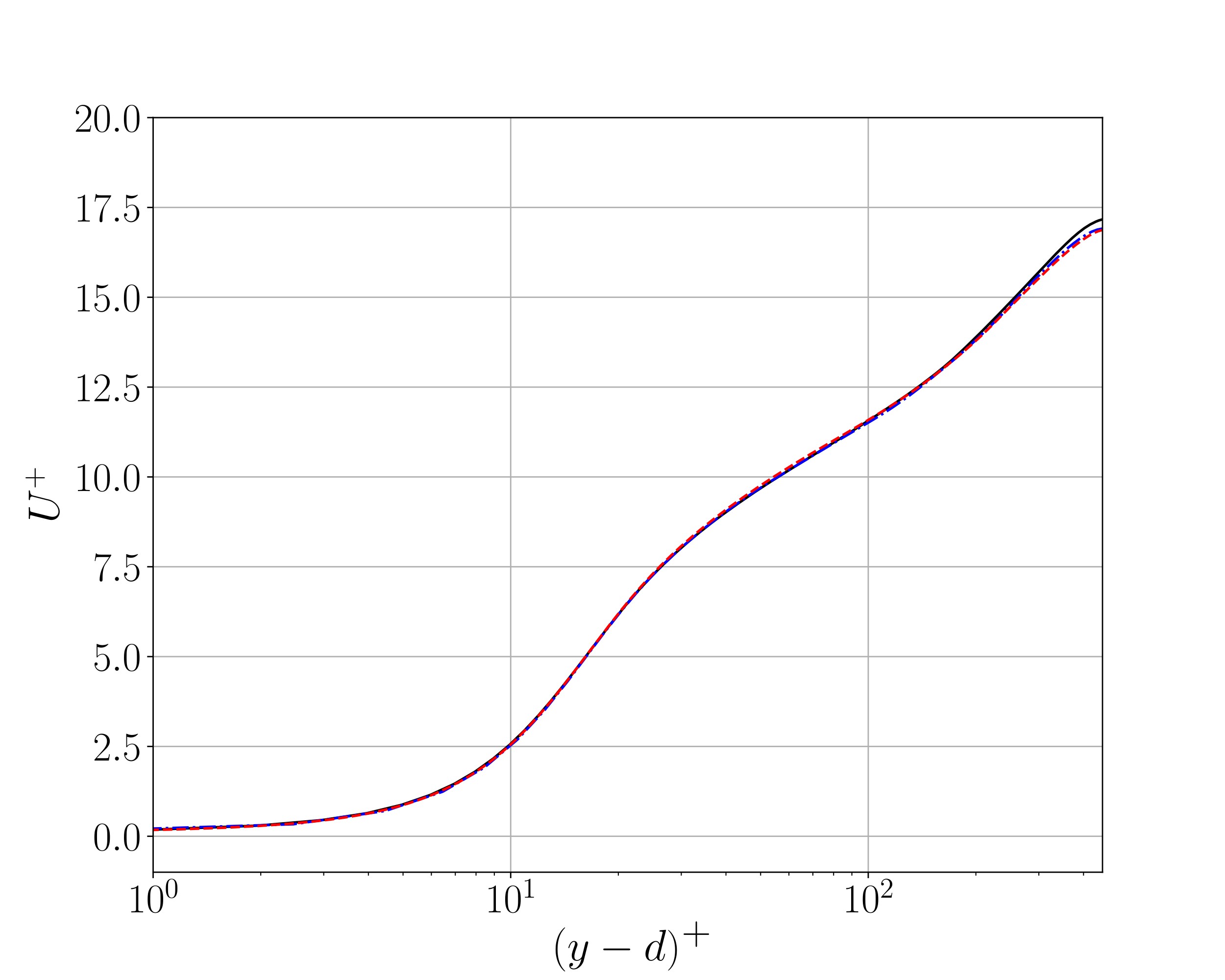}
\caption{Mean streamwise velocity profile of each different grid resolution: 
\protect\blackline, $\text{Base grid}$; 
\protect\DNSblueline, $\text{Coarse grid}$; 
\protect\DNSredline, $\text{Fine grid}$.}
\label{fig:GCS}
\end{figure}
The grid numbers and resolutions utilized in this grid convergence test are summarized in Table~\ref{tab:GCS}. Fig.~\ref{fig:GCS} illustrates the mean velocity profile in the $x$-direction for each grid resolution. When the fine grid is considered as the ground truth, the absolute percentage errors at $y_{ref}$ are $0.969\%$ for the base grid and $0.808\%$ for the coarse grid. A more diverse validation of the DNS used in this study, such as Reynolds shear stress, is included in another paper by the authors~\cite{shi2024drag}.

\section{3D Visualization of DNS drag maps and CNN feature maps} \label{app:3d_vis}
\begin{figure*}
    \centering
    \includegraphics[width=1.3\textwidth, angle=-90]{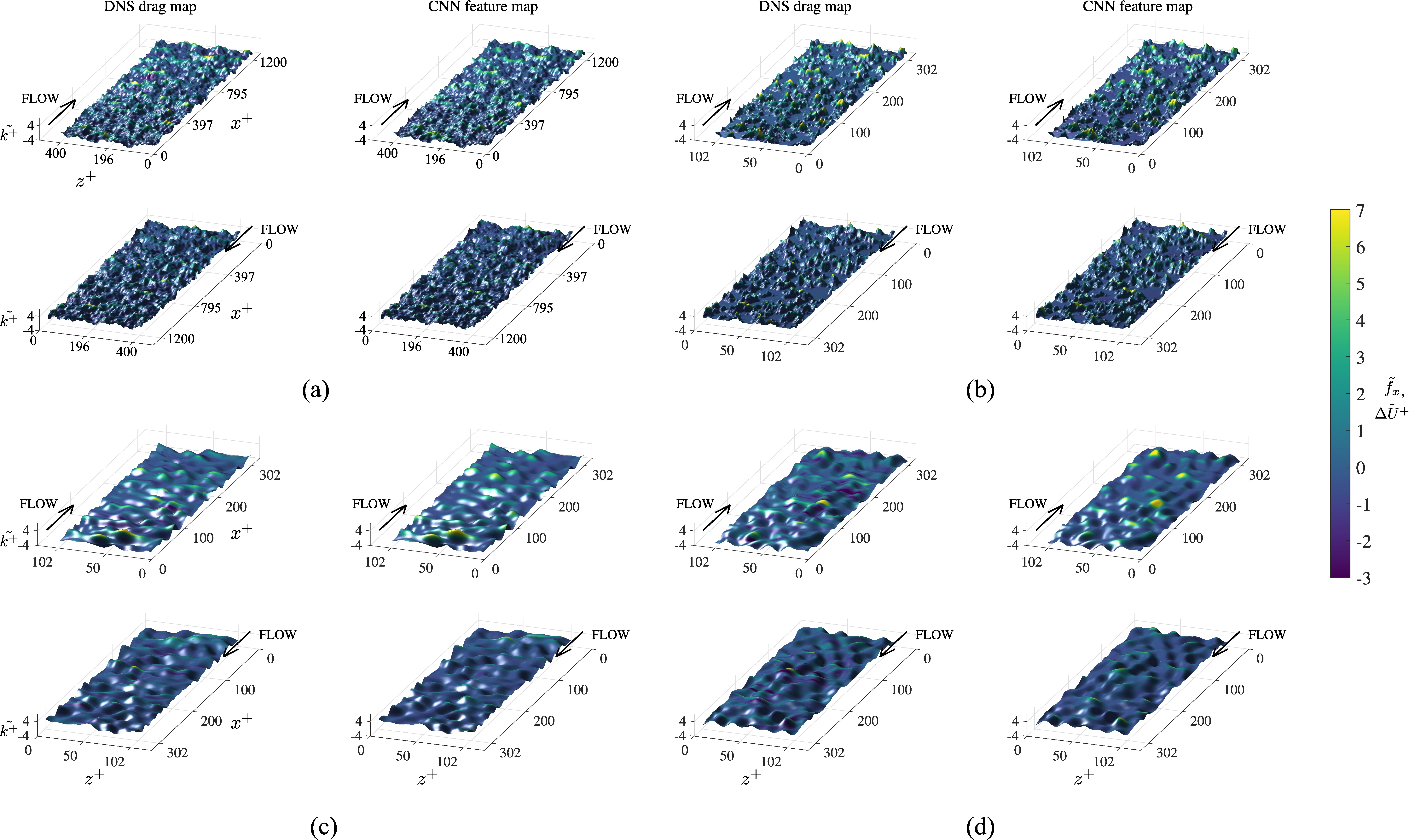}
    \caption{\label{fig:3d_maps}Visualization of the DNS drag map and CNN feature map in 3D, with panels (a) to (d) corresponding to surfaces $S_{Gauss,1}$, $S_{pos,1}$, $S_{ES_x,1}$, $S_{ES_z,1}$, respectively.}
    \label{fig:3d}
\end{figure*}

\begin{figure*}
    \centering
    \includegraphics[width=1\textwidth]{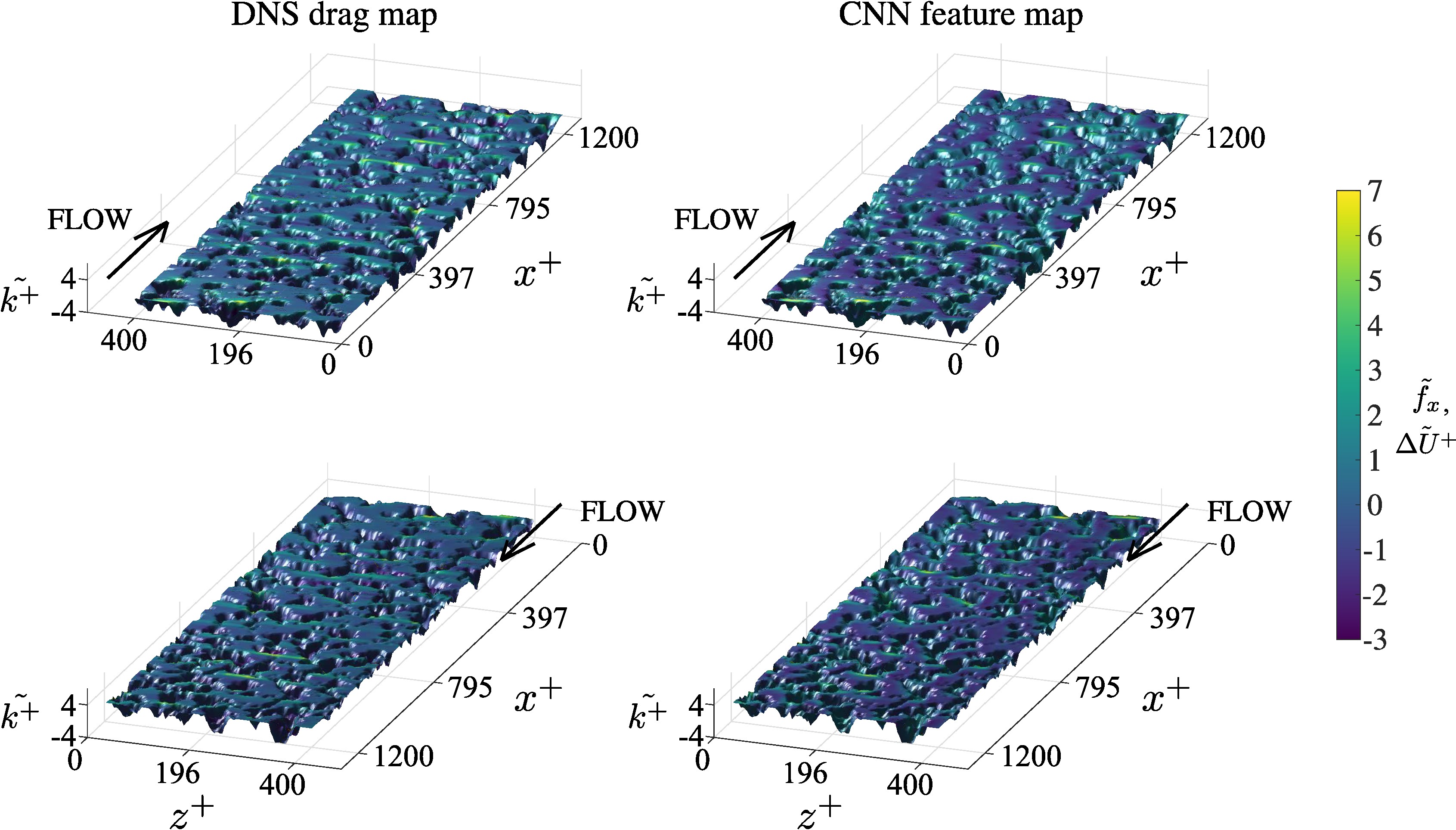}
    \caption{Visualization of the DNS drag and CNN feature maps on a topographical map in 3D for $S_{neg,1}$.}
    \label{fig:neg_skewness_3d}
\end{figure*}

To further analyze and understand the patterns appearing on the maps, we visualized the DNS drag maps and CNN feature maps in three dimensions. Fig.~\ref{fig:3d} is a 3D visualization of the CNN feature map and DNS drag map from the perspective of the flow direction and the opposite direction on the surfaces $S_{Gauss,1}$, $S_{pos,1}$, $S_{ES_x,1}$, and $S_{ES_z,1}$. Similarly, Fig.~\ref{fig:neg_skewness_3d} is a 3D visualization of $S_{neg,1}$ viewed from the flow direction and the opposite direction.

\section{Two-dimensional premultiplied PSD} \label{app:2d_psd}
\begin{figure*}
\centering
\includegraphics[width=1\textwidth]{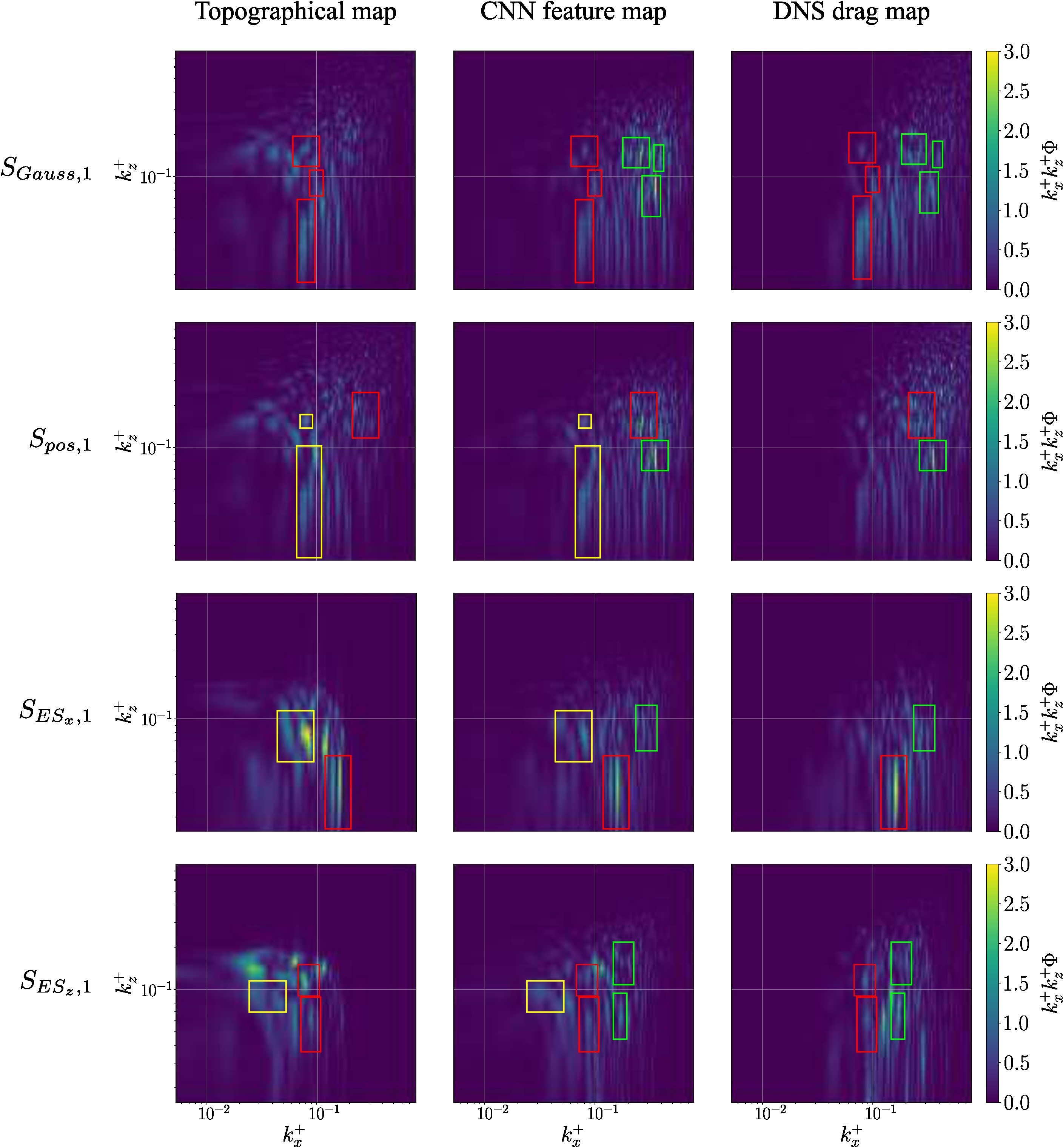}
\caption{$k_x^+ k_z^+ \Phi$ of the topographical-, CNN feature-, and DNS drag maps, regarding $S_{Gauss,1}$, $S_{pos,1}$, $S_{ES_x,1}$, and $S_{ES_z,1}$. The marked box areas denote regions of similar high-intensity distributions between the topographical maps and the CNN feature maps (yellow boxes), between the DNS drag maps and the CNN feature maps (green boxes), and across all three types of maps (red boxes).}
\label{fig:2d_PSD}
\end{figure*}

We analyzed the two-dimensional premultiplied PSD ($k_x^+ k_z^+ \Phi$), which are shown in Fig.~\ref{fig:2d_PSD}.
\begin{ruledtabular}
\begin{table}
\begin{center}
\begin{tabular}{lcccc}
& $S_{Gauss}$ & $S_{pos}$ & $S_{ES_x}$ & $S_{ES_z}$\\
\hline
CNN - DNS & 0.424 & 0.345 & 0.784 & 0.753 \\
CNN - Topo & 0.350 & 0.322 & 0.644 & 0.584\\
CNN - DNS & 0.070 & 0.082 & 0.046 & 0.051 \\
CNN - Topo & 0.080 & 0.078 & 0.052 & 0.058 \\
\end{tabular}
\caption{Comparison of $k_x^+ k_z^+ \Phi$ using the average SSIM (top) and RMSE (bottom) of three samples.}
\label{tab:combined_2D_PSD}
\end{center}
\end{table}
\end{ruledtabular}
The similarities of the patterns in the samples' $k_x^+ k_z^+ \Phi$ were evaluated using the SSIM and RMSE, as presented in Table~\ref{tab:combined_2D_PSD}. This table indicates that, for all the samples, the SSIM between the $k_x^+ k_z^+ \Phi$ of the DNS drag map and that of the CNN feature map is higher than that between the $k_x^+ k_z^+ \Phi$ of the CNN feature map and that of the topographical map, except for $S_{pos,3}$. In terms of the RMSE, the values are generally lower between the CNN feature maps and the DNS drag maps, except for $S_{pos,1}$, $S_{pos,2}$, and $S_{ES_x,2}$. Both the RMSE and SSIM metrics corroborate the findings from the $k_x^+ \Phi$ analysis, demonstrating a resemblance between the CNN feature maps and the DNS drag maps. Fig.~\ref{fig:2d_PSD} not only confirms the congruence of the CNN feature maps with the DNS drag maps in terms of a high-intensity distribution but also highlights significant similarities in this distribution with the topographical maps. For instance, the marked boxes in $S_{ES_x,1}$ of Fig.~\ref{fig:2d_PSD} show common distribution patterns between the DNS drag and the CNN feature maps, and between the topographical and the CNN feature maps, also capturing distributions common to all the maps. This underscores the capability of the CNN model to extract dominant spatial features in surface topography and identify the essential scales of spatial patterns for predicting $\Delta U^+$ over rough surfaces.
\begin{figure*}
\centering
\includegraphics[width=1\textwidth]{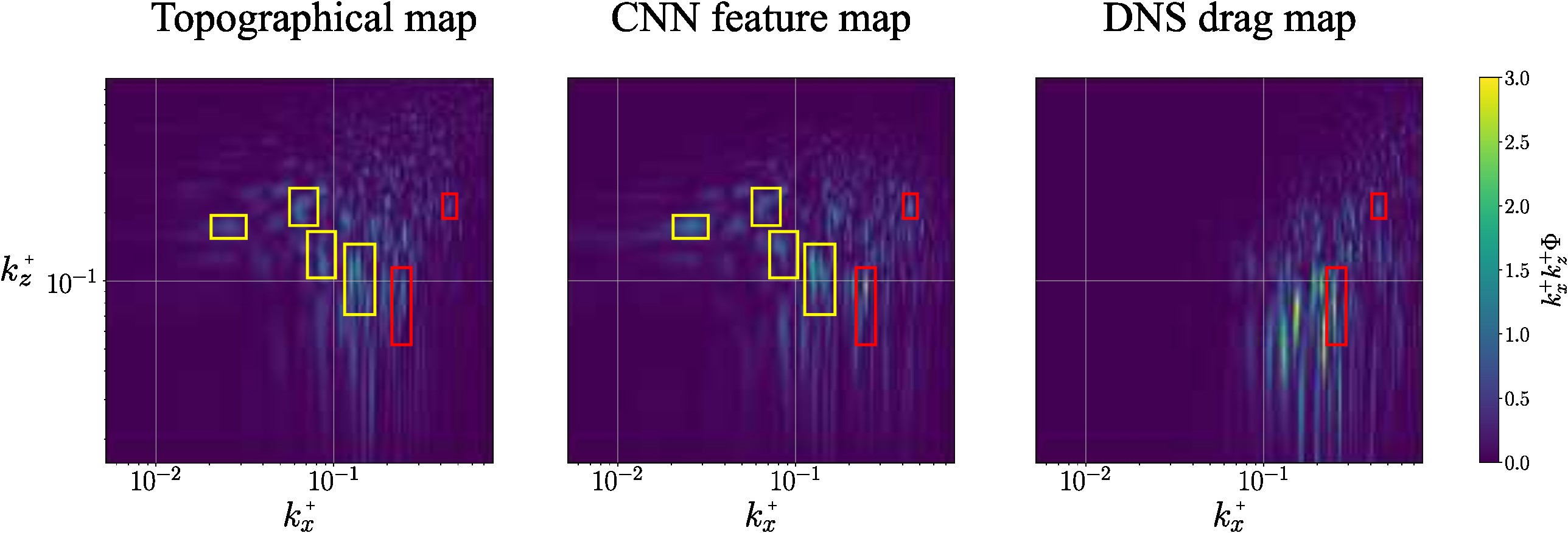}
\caption{$k_x^+ k_z^+ \Phi$ of the topographical maps, CNN feature maps, and DNS drag maps, regarding $S_{neg,1}$. The marked box areas denote regions of similar high-intensity distributions between the topographical maps and the CNN feature maps (yellow boxes), between the DNS drag maps and the CNN feature maps (green boxes), and across all three types of maps (red boxes).} 
\label{fig:2d_PSD_neg}
\end{figure*}
Unlike surface types in Fig.~\ref{fig:2d_PSD}, the scale of spatial patterns in the CNN feature map of $S_{neg}$ more closely resembles that in the topographical map, diverging from the distributions observed in the DNS drag map. This congruence is evident in the $k_x^+ k_z^+ \Phi$ (see Fig.~\ref{fig:2d_PSD_neg}). This results suggest that the CNN model did not accurately capture the spatial patterns for $S_{neg}$, unlike other previously analyzed surface types.

\section{\label{app:architecture}Architectural features of the CNN model}
The architecture of our model was designed to include several important features: 
\begin{enumerate}
    \item It utilizes a residual network (ResNet) framework, as proposed by~\citet{he2016deep}, which is extensively employed to address the vanishing gradient problem. The ResNet architecture incorporates skip connections, which create direct pathways to previous layers, effectively addressing the vanishing gradient problem. This problem involves the diminishing gradients of the loss function during the training of deep neural networks, leading to minimal parameter updates. The skip connections mitigate this issue by ensuring a consistent and effective flow of gradients throughout the network. We employed ResNet to improve our model's predictive performance.
    
    \item It adopts periodic boundary padding to reflect the periodic boundary condition used in DNS, ensuring consistency between the simulation and the model input.
    
    \item The architecture uses GAP to create a feature map that effectively emphasizes critical areas on the rough surface that are important for predicting $\Delta U^+$. GAP, as outlined by \citet{zhou2016learning}, is crucial for generating a two-dimensional feature map that highlights significant regions in the data, thereby aiding the decision-making processes of the CNN. Additionally, using GAP in the final layer not only acts as a regularizer but also reduces the use of Fully Connected Networks (FCN), simplifying the model.
    
    \item It features a parallel structure with various kernel sizes, inspired by the Inception module proposed by \citet{szegedy2015going}. The Inception module enhances a model's ability to capture multiscale features. To better capture the multiscale spatial patterns of rough surfaces and improve drag prediction, we integrated the Inception module.
\end{enumerate}

We conducted a sensitivity study on kernel size to examine the changes in the CNN feature map and predictive performance, particularly regarding the utilization of the Inception module. Since CNN kernels act as wavenumber filters that learn the spatial scale, we tested the impact of utilizing both single or multiple kernel sizes. The test cases are as follows: (a) kernel size=3, (b) kernel size=5, (c) kernel size=7, (d) kernel size=9, (e) kernel size=11, and (f) our model using kernel size=3, 5, 7, 9, and 11.
\begin{figure*}
    \centering
    \includegraphics[width=1\linewidth]{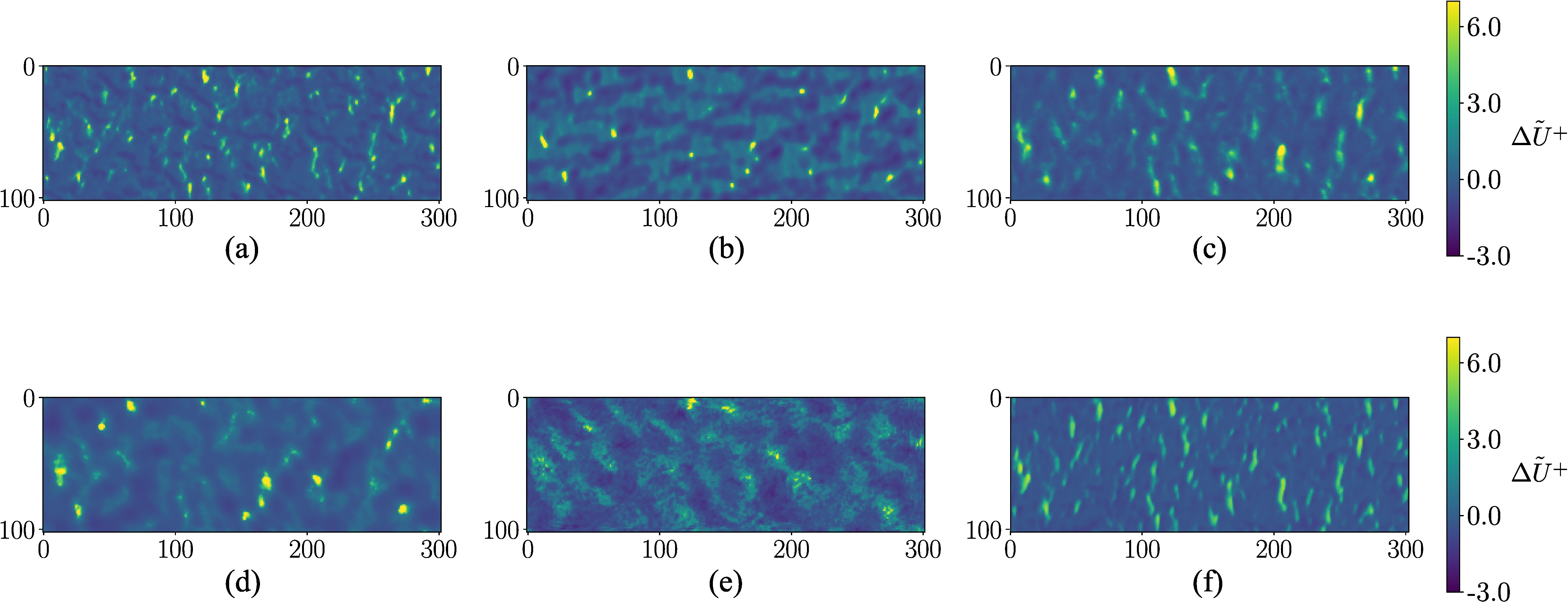}
    \caption{The CNN feature maps of $S_{Gauss,1}$ generated from the CNN models used in a kernel size sensitivity analysis. The kernel sizes used for each model are as follows: (a) kernel size=3, (b) kernel size=5 (c) kernel size=7 (d) kernel size=9, (e) kernel size=11, and (f) 3, 5, 7, 9, 11 (Our model used in this study).}
    \label{fig:ks}
\end{figure*}
\begin{ruledtabular}
\begin{table}
    \centering
    \begin{tabular}{lcccccc}
       & Our model & kernel size=3 & kernel size=5 & kernel size=7 & kernel size=9 & kernel size=11 \\
       \hline
        MAE& 0.105 & 0.138 & 0.120 & 0.110 & 0.104 & 0.140 \\
        $R^2$& 0.996 & 0.995 & 0.996 & 0.996 & 0.997 & 0.995 
    \end{tabular}
    \caption{MAE and $R^2$ for sample surfaces from the models with different kernel sizes. Our model utilized all kernel sizes (3, 5, 7, 9, 11)}
    \label{tab:ks_acc}
\end{table}
\end{ruledtabular}
\begin{ruledtabular}
\begin{table}
    \centering
    \begin{tabular}{lcccccc}
       & Our model & kernel size=3 & kernel size=5 & kernel size=7 & kernel size=9 & kernel size=11 \\
       \hline
        $S_{Gauss}$& 0.490 & 0.473 & 0.240 & 0.454 & 0.438 & 0.230\\
        $S_{pos}$ & 0.594 & 0.564 & 0.325 & 0.573 & 0.466 & 0.281 \\
        $S_{neg}$ & 0.058 & 0.094 & 0.101 & 0.084 & 0.105 & 0.050 \\
        $S_{ES_x}$& 0.474 &  0.463 & 0.203 & 0.441 & 0.414 & 0.250\\
        $S_{ES_z}$& 0.410 & 0.400 &  0.224 & 0.363 & 0.376 & 0.255\\
    \end{tabular}
    \caption{Mean SSIM for sample surfaces from the models with different kernel sizes. Our model utilized all kernel sizes (3, 5, 7, 9, 11)}
    \label{tab:ks_ssim}
\end{table}
\end{ruledtabular}
Fig.~\ref{fig:ks} visualizes the CNN feature map for sample 1 of the $S_{Gauss}$ surface used in the sensitivity test for the models. According to the sensitivity study, the predictive performance of all models was similar (see Table~\ref{tab:ks_acc}). However, among all the cases except for $S_{neg}$, the Full model using kernel sizes 3, 5, 7, 9, and 11, which is the same as the model used in the manuscript, showed the highest similarity between the CNN feature map and the DNS drag map based on SSIM (see Table~\ref{tab:ks_ssim}). In conclusion, our model, which utilizes a combination of various kernel sizes, effectively captures the spatial wavenumber information of random roughness elements at different scales to learn the physics related to drag induced by rough surfaces.

\section{\label{app:data}Data preprocessing} 
\subsection{Data augmentation}
Hydrodynamically smooth surfaces, denoted as $S_{smooth}$, were included to enhance the diversity of the dataset. As $U_R^+$ in Eq.~\ref{eqn:U^+} equals $U_S^+$ for these surfaces, the resulting $\Delta U^+$ is consistently zero. This allowed us to create a smooth surface dataset without using additional DNS. 

Furthermore, the dataset was expanded by reflecting surfaces along the $x$-axis. This mirrored surface retains its original $\Delta U^+$ value, making it suitable for augmentation. This strategy effectively doubles the dataset size without requiring additional computations. 

\subsection{Data partitioning}
\begin{figure*}
\centerline{
\includegraphics[width=\textwidth]{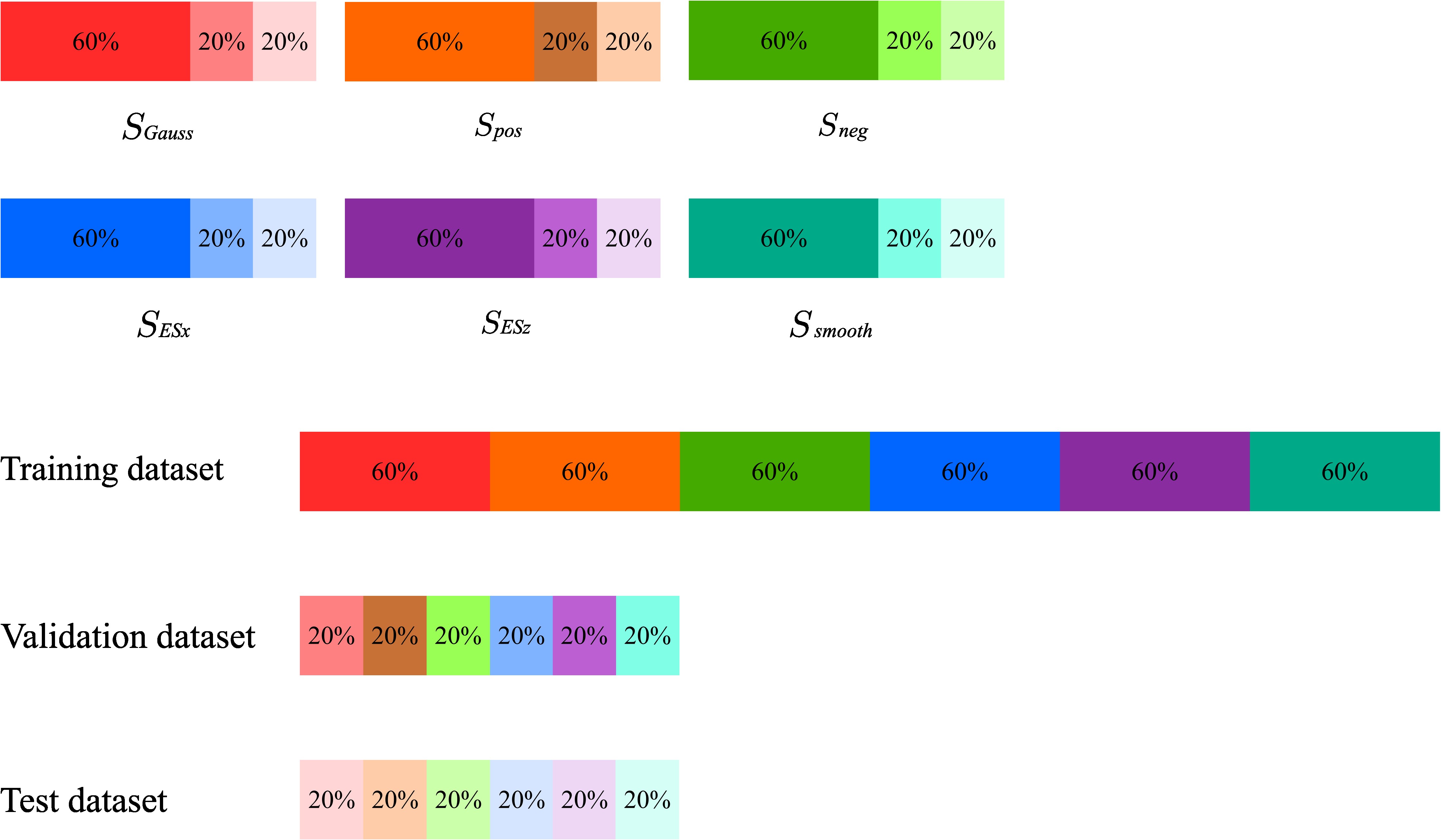}}
\caption{Partitioning of dataset into training, validation, and test subsets.}
\label{fig:data_split}
\end{figure*}

The expanded dataset of topographical maps was divided into training, validation, and test sets in the proportions of 60\%, 20\%, and 20\%, respectively, as illustrated in Fig.~\ref{fig:data_split}.
To optimize model training and minimize biases, the rough surfaces and their corresponding $\Delta U^+$ values were randomly shuffled. This procedure ensured an even distribution of data throughout the CNN training phase. All the data were then standardized using Eq.~\ref{eqn:std_scaler}, based on the mean and variance of the training dataset.

\section{\label{app:opt}Hyperparameter optimization}
The optimization of hyperparameters is essential for improving the performance of ANNs. A thorough analysis of these parameters was performed to refine the CNN model. Bayesian optimization (BO) was employed for this purpose. BO uses a probabilistic model to predict the performance of the objective function, enabling an efficient exploration of the hyperparameter space of machine learning model~\cite{snoek2012practical}. This method is particularly beneficial in scenarios involving high-dimensional optimization, where exhaustive searches are computationally infeasible. In this study, the hyperparameters optimized for the CNN model through BO are $N_b = 3$ and $N_f = 48$, where $N_f$ represents the number of filters, and $N_b$ denotes the number of ResNet blocks.

\bibliography{main}
\end{document}